\documentclass[11pt]{article}

\usepackage[margin=1in]{geometry}
\usepackage{setspace}
\usepackage[T1]{fontenc}
\usepackage[utf8]{inputenc}
\usepackage{lmodern}
\usepackage{microtype}
\usepackage{sgame}

\usepackage{amsmath,amssymb,amsthm,mathtools,bm}

\usepackage{graphicx}
\usepackage{booktabs}
\usepackage{caption}
\usepackage{subcaption}
\usepackage{rotating} 

\usepackage{enumitem}

\usepackage[round,authoryear]{natbib}
\usepackage{hyperref}
\hypersetup{
  colorlinks=true,
  linkcolor=blue,
  citecolor=blue,
  urlcolor=blue
}
\usepackage{cleveref} 

\numberwithin{equation}{section}
\newtheorem{theorem}{Theorem}[section]
\newtheorem{proposition}[theorem]{Proposition}
\newtheorem{lemma}[theorem]{Lemma}
\newtheorem{corollary}[theorem]{Corollary}
\newtheorem{assumption}[theorem]{Assumption}
\theoremstyle{definition}
\newtheorem{definition}[theorem]{Definition}
\theoremstyle{remark}

\newtheorem{lemmaapp}{Lemma A}

\usepackage{mathtools}
\usepackage[usenames,dvipsnames]{xcolor}
\usepackage[toc,page]{appendix}
\usepackage{graphicx}
\usepackage{pgfplots} 
\pgfplotsset{compat=1.18} 
\usepackage{enumitem}
\usepackage{scrextend}
\usepackage{epstopdf}
\usepackage{cancel}
\usepackage{comment}
\usepackage{tikz}
\usepackage{bbm}
\usetikzlibrary{patterns}
\usetikzlibrary{decorations.pathreplacing}
\usepackage{sgame}
\usepackage{color}
\usepackage{indentfirst}
\usepackage{natbib}
\setcitestyle{authoryear,round}
\usepackage{footmisc}
\usepackage{wasysym}
\usepackage[capposition=bottom]{floatrow}

\newcommand{\x}{\mathbf{x}}
\newcommand{\p}{\mathbf{p}}

\newcommand{\dd}{\mathrm{d}}
\newcommand{\indep}{\perp \!\!\! \perp}

\title{Belief Diversity and Cooperation}
\author{Georgy Lukyanov\thanks{École Polytechnique, 5 Av. Le Chatelier, 91120 Palaiseau, France}
\and
David Li\thanks{International College of Economics and Finance, 11 Pokrovsky Boulevard, building T, Moscow, 109028, Russia}}
\date{}

\begin{document}
\maketitle

\begin{abstract}
This paper studies a two-player game in which the players face uncertainty regarding the nature of their partner. In this variation of the standard Prisoner's Dilemma, players may encounter an 'honest' type who always cooperates. Mistreating such a player imposes a moral cost on the defector. This situation creates a trade-off, resolved in favor of cooperation if the player's trust level, or belief in their partner's honesty, is sufficiently high. We investigate whether an environment where players have explicit beliefs about each other's honesty is more or less conducive to cooperation, compared to a scenario where players are entirely uncertain about their partner's beliefs. We establish that belief diversity hampers cooperation in environments where the level of trust is relatively low and boosts cooperation in environments with a high level of trust.
\end{abstract}



\medskip

\noindent
\begingroup
\setlength{\fboxsep}{8pt}%
\setlength{\fboxrule}{0.4pt}%
\fbox{%
  \begin{minipage}{0.97\textwidth}
  \small
  \textbf{Author's accepted manuscript (postprint)} of:
  Lukyanov, G., \& Li, D. (2025). \emph{Belief diversity and cooperation}.
  \textit{Journal of Economic Behavior \& Organization}, 229, 106815.
  \textbf{Version of Record:}
  \href{https://doi.org/10.1016/j.jebo.2024.106815}{10.1016/j.jebo.2024.106815}.
  \textbf{License for this manuscript:} CC BY-NC-ND 4.0.
  \end{minipage}%
}
\endgroup

\bigskip

\section{Introduction}

It is assumed that altruistic actions stem from either the expectation of future rewards or the avoidance of social ostracism. Yet, tolerance of opportunistic behavior in certain contexts can significantly diminish cooperative endeavors. Often, the opportunistic attitude falls short of fostering cooperation, making the role of individuals who adhere to deontological ethics, those who inherently 'do the right thing,' crucial for the functioning of institutions.\footnote{For example, \cite{Strulovici2020} shows that there is no effective method to motivate researchers dealing with historical data to ascertain accurate facts unless they are inherently driven by a commitment to truthfulness for its own merit.} On the other hand, cooperation can partly be restored if the rationality (narrowly defined as maximizing one's own payoff) of the player's partner is questioned.

Even for individuals who do not strictly adhere to ethical principles, their behavior is greatly shaped by the anticipation of engaging with a reliable counterpart. In routine interactions, it's common to presume that the involved parties share a mutual belief in the probability of encountering an honest partner. However, when the individuals hail from diverse backgrounds, it's reasonable to expect a degree of uncertainty regarding each other's trustworthiness. In some situations, an individual may be highly optimistic about their partner's honesty, yet remain unaware of their partner's belief regarding their own honesty. Compare this to the scenario where both parties are relatively pessimistic but have common knowledge of each other's beliefs. In which of these two cases would one expect greater cooperation?

Addressing this issue, we consider a two-player Prisoner's Dilemma with incomplete information. Each player can be of two types, `honest' and `strategic.' An `honest' type is the one who plays the committed action (cooperation), whereas the strategic player faces the choice between cooperation and defection. While defecting from a strategic partner who cooperates brings a net gain, if it happens that the strategic player encounters an honest partner, then defection inflicts on him a moral cost. To simplify exposition and confine ourselves to pure strategies, we also assume that each player has private information concerning the loss that he would suffer from cooperating with a partner who chooses to defect.

This creates a trade-off for the strategic player. His ultimate decision will depend on his \emph{belief} concerning the likelihood of being matched with an honest partner. To that end, we focus on two scenarios: one in which the two players hold identical, commonly known beliefs, and the other in which the players hold diverse beliefs regarding their partner's honesty. The main question addressed in this paper is: How does the diversity in these beliefs impact the level of cooperation?

We first establish that in any equilibrium, the player's strategy can be described by a simple cutoff rule (Lemma \ref{lemma:threshold}): for any belief about his partner's honesty, the strategic player cooperates if and only if his loss from suffering defection falls below a threshold. In section \ref{subs:common}, we proceed to the setting in which the two players hold identical, commonly known beliefs about the likelihood of facing an honest partner. We show that there always exists a symmetric equilibrium and establish the conditions of the primitives when this equilibrium is unique (Proposition \ref{prop:threshold}). Furthermore, we show that the cutoff is increasing in the player's level of belief of meeting an honest partner, which is intuitive: if I place a high probability that my partner is honest (and thus, will certainly cooperate), then I cooperate for a larger range of losses from defection.

Section \ref{subs:diverse} proceeds to analyze the diverse beliefs case---the situation in which the player does not know the belief held by his partner. In that situation, we show that in any symmetric equilibrium, the player's strategy can be described by the cutoff rule on his belief (Lemma \ref{lem:divthresh}): for any loss from the partner's defection, the strategic player cooperates if and only if his belief that his partner is honest is above a threshold. Then we establish that for any non-degenerate distribution of beliefs, there exists a unique threshold function (Proposition \ref{main:diverse}) that is increasing: the critical value for the belief above which the player cooperates is larger (and hence, cooperation is less likely), the higher is his loss from his partner's defection.

The equilibrium characterization in Section \ref{sec:equilibrium} paves our way to our main results concerning the probability of cooperation. Section \ref{sec:cooperation} takes the specific case in which both the losses from cooperating with a defecting partner and beliefs are drawn from the uniform distribution on $[0,1]$. This case allows us to obtain closed-form expressions (Proposition \ref{prop:heterogeneous}) for the threshold rules under the two cases---with common and diverse beliefs, respectively. The results, illustrated on Figures \ref{fig:twothresh}(a) and (b) suggest that the common knowledge of beliefs makes the players more likely to cooperate when their beliefs are low. However, when their beliefs are high, belief diversity \emph{helps} cooperation.

Intuitively, this happens for the following reason: For low beliefs, both players are almost certain that they are facing a strategic dishonest partner, and in that situation the structure of the game is the same as in the classical \emph{Prisoner's Dilemma}. In particular, the players' strategies (as described by their threshold functions) are \emph{strategic substitutes}: a reduction in the threshold of my partner raises the chance that he will cooperate, which in turn makes me \emph{less eager to cooperate}. On the other hand, for higher levels of beliefs, the game structure starts to resemble the \emph{coordination game}, in which a higher willingness of my partner to cooperate makes me more willing to cooperate myself. (This reasoning is illustrated on different panels of Figure \ref{fig:eqthresh}). Hence, belief diversity pushes the players to defect when beliefs are low and to cooperate when beliefs are high.

Based on that, we analyze the likelihood of cooperation, both from an \emph{ex post} and an \emph{ex ante} perspective. From the viewpoint of someone who can observe the players' beliefs, we claim that belief diversity raises the likelihood of cooperation whenever those beliefs are higher than a certain value (Proposition \ref{prop:dagger}). The graphical analysis on Figure \ref{fig:pimg} reveals that this value is low when the benefits to defection are small and the moral cost of defecting an honest partner is large, suggesting the instances in which belief diversity is especially valuable in terms of promoting cooperation.

Finally, we compute the probability of cooperation from the viewpoint of someone who cannot observe the players' beliefs. This case allows us to compute and compare the unconditional probabilities that the strategic player chooses to cooperate under the two scenarios, with common and diverse beliefs, respectively. Proposition \ref{prop:exante} derives the closed-form expression for those probabilities, and Figure \ref{fig:exante} reveals that belief diversity raises the likelihood of cooperation when the moral costs are not too large.

\subsection{Literature review}

There is a large body of literature discussing the factors that facilitate cooperation in the Prisoner's Dilemma. Broadly speaking, these can be classified into one of three groups: (i) repeated interaction, (ii) preplay communication, and (iii) uncertainty about others' motives.

Our treatment of `honesty' follows \cite{Frank1987}, who considers the evolutionary model in which the population of agents includes some who `always do the right thing,' while others may `act opportunistically.' However, the focus of his paper was on the expression of certain social cues (such as uncontrolled blushing when telling a lie) that may be used to signal one's honesty. By contrast, in our model, the player's desire to cooperate depends on his belief that his partner is honest, since mistreating such a partner involves a cost.

One way to formalize this moral cost may be through some form of guilt, which places our setup in the context of the literature on lying and guilt aversion, as formalized by \cite{Gneezy2005} and \cite{BattigalliDufwenberg2007}. In \cite{Baheletal2022}, players can exchange messages prior to playing the Prisoner's Dilemma, and the violation of the previously given promise to cooperate is costly. They show that as the players' value of honesty increases, it might make them less willing to cooperate. In our setup, we do not explicitly model preplay communication; however, the moral cost we introduce could be interpreted as a feeling of guilt from betraying our partner's trust. From this perspective, we consider the initial stage in which the players exchange promises, and both players might be uncertain whether the partner interprets the initial (presumably, informal) agreement as binding. If so, then the violation of this agreement would be costly, for example, if the player whose trust was betrayed takes legal action.

There exists a rich experimental literature testing guilt aversion. \cite{Ellingsenetal2010} measure individuals' utility loss from disappointing someone (the analogue of the moral cost of defecting from an honest partner in our setup) in the dictator game and the trust game. \cite{Vanberg2008} conducts a series of dictator games to test whether honest behavior is driven more by people's promise-keeping preferences than by the fear of disappointing their partners. \cite{EllingsenJohannesson2004} conduct an experiment to show how a player's promises can resolve the hold-up problem.

While in those settings, the players have private information about their preferences, the \emph{distribution} of preferences is commonly known. By contrast, at the heart of our work is belief diversity---whereby the players do not know how likely it is that they will encounter an honest partner---and its impact on their decisions to cooperate. Although our paper focuses purely on theory, in Section \ref{sec:experdes}, we offer a preliminary outline of how the implications of our setup could be tested experimentally.

The impact of belief diversity on players' behavior bears some resemblance to that in the global games literature, initially proposed by \cite{CarlssonVanDamme1993} and subsequently developed in a series of papers by \cite{MorrisShin1998,MorrisShin2003}. This literature considers a class of incomplete-information coordination games where the players face minor, privately observed payoff perturbations, enabling the formulation of their strategies as simple threshold rules. We adopt a similar approach by introducing privately observed losses from cooperating with a defecting player.

\cite{KetsSandroni2020} examine the impact of cultural diversity on individuals' behavior in coordination games with Pareto-ranked equilibria. Their concept of `cultural distance' effectively captures the extent to which different groups share cognition: the members of each group receive \emph{impulses} to act in a certain way, and cultural proximity reflects how these inclinations can inform similar impulses in another group. Their main normative conclusions are that cultural diversity is beneficial as it enables societies to avoid coordinating on Pareto-inferior outcomes, but it also increases the risk of miscoordination. In this paper, we focus on the diversity arising from beliefs about encountering an honest partner. Accordingly, in our game, the payoff structure is modified such that the strength of the coordination motive intensifies with those beliefs.

In a multi-period context, the cost of mistreating an honest partner might be represented by the reputational damage resulting from ostracism, as discussed by \cite{HirshleiferRasmusen1989}. The traditional belief that high future valuations foster cooperation in long-term relationships was challenged by \cite{SkaperdasSyropoulos1996}, who showed that an extended `shadow of the future' might actually reduce cooperation. While we provide a brief discussion (see Section \ref{subs:repext}) on how the moral cost can emerge endogenously in a repeated framework where players may develop a reputation for honesty, our main analysis is situated in a static setting where the motives of honest players are considered a black box, and strategic players' behavior is influenced by their beliefs about encountering honest partners and the associated costs of mistreating them.

\indent

The remainder of the paper proceeds as follows:

Section \ref{sec:model} provides a detailed description of the modified prisoner's dilemma. Section \ref{sec:equilibrium} offers a characterisation of the equilibrium for two cases: common and heterogeneous backgrounds, based on Propositions \ref{prop:threshold} and \ref{prop:heterogeneous}. Section \ref{sec:cooperation} deepens the analysis by focusing on a specific parametric case. In this section, we discuss comparative statics results regarding the probability of cooperation. Section \ref{sec:applications} extends the discussion to potential applications and ramifications of our findings. Section \ref{sec:conclusion} summarises our conclusions and provides direction for future research.

\section{\label{sec:model}Model}

There are two agents, $i = 1, 2$, who are randomly matched to play a game. Each agent can either choose to cooperate ($C$) or defect ($D$). Each agent can be of two types, `honest' and `strategic.' By definition, the `honest' type is the one who always chooses $C$.

Player $i$ holds a belief $\pi_i$ that his partner, player $j\neq i$, is `honest.' With complementary probability $(1-\pi_i)$, player $i$ thinks that his partner is `strategic.'

If the two `strategic' players are matched, they play the game described by the following matrix:
\begin{center}
\begin{game}{2}{2}[Agent~$i$][Agent~$i$'s partner]
          	&$D$	 			&$C$ \\
$D$		&$0,0$	 			&$b,-\ell_j$  \\
$C$		&$-\ell_i,b$	 			&$1,1$
  \end{game}
\end{center}

The parameter $b>1$ denotes the benefit from defection and $\ell_i>0$ ($i=1,2$) represents the cost from cooperating with a defecting partner. We assume that $\ell_i$ is initially drawn from a cdf $F(\cdot)$ on $[0,\overline{\ell}]$ for some $\overline{\ell}>0$, with the corresponding density denoted by $f(\cdot)>0$. Each agent $i$ privately observes his value of $\ell_i$.

With probability $\pi_i$, the player $i$ is matched with an honest partner who always cooperates. If the strategic player $i$ defects playing against an `honest' type, he incurs a cost $m>0$, so that his net benefit from defection is equal to $(b-m)$. The corresponding payoff matrix describing the game between (strategic) player $i$ and (honest) player $j$ is described by the following matrix:

\begin{center}
\begin{game}{2}{1}[Agent~$i$][Agent~$j$]
          			&$C$ \\
$D$				&$b-m,-\ell_j$  \\
$C$				&$1,1$
  \end{game}
\end{center}

The parameters $b$ and $m$ are commonly known.

To make things non-trivial, we assume that $m$ is large enough, so that the `strategic' player would prefer to cooperate with his partner, if he were certain that the partner is `honest'.

\begin{assumption}
\label{ass:par}
The parameters $b$ and $m$ satisfy $m>b-1$.
\end{assumption}

To parametrize the agents' knowledge about their environment by their beliefs about encountering the `honest' type, we consider two distinct cases: the case with \emph{common beliefs} and the case \emph{diverse beliefs}.

\begin{definition}
In case of \emph{common beliefs}, the two agents hold identical beliefs $\p_1=\pi_2=\pi$, where $\pi\in[0,1]$ is commonly known.

In case of \emph{diverse beliefs}, from agent $i$'s perspective, his partner's belief, $\pi_j$, follows a distribution $\pi_j\sim_{i}G(\cdot)$, with continuous density $g(\cdot)>0$.
\end{definition}

We define an agent $i$'s (mixed) strategy as a mapping from his information set into the probability distribution over his actions. Formally,
\begin{equation}
\label{eq:strategy}
\sigma_i:[0,\overline{\ell}]\times[0,1]\to\Delta A.
\end{equation}

We adopt the Bayesian Nash equilibrium as the solution concept.

\begin{definition}
An equilibrium is a pair of strategies $(\sigma_1,\sigma_2)$ as specified in \eqref{eq:strategy}, such that each agent maximizes his expected payoff: for every $i$ and every $(\ell,\pi)\in[0,\overline{\ell}]\times[0,1]$,
\begin{equation}
\sigma_{i}(\ell,\pi)\in\underset{s}{\arg\max}\int_0^1\int_0^{\overline{\ell}}u_i(s,\sigma_j(\ell_j,\pi_j);\ell,\pi)\dd F(\ell_j)\dd G(\pi_j).
\end{equation}
\end{definition}

In the next section, we establish the existence of equilibrium for the two cases of \emph{common beliefs} and \emph{diverse beliefs}. Since we assume that the players are ex ante identical, we will constrain our attention to \emph{symmetric strategies}, eliminating the need for subscripts from now on.

\section{\label{sec:equilibrium}Equilibrium Characterization}

To pave our way for equilibrium characterization, we first establish one useful result (Lemma \ref{lemma:threshold}), which shows that the player's strategy can be described by a simple \emph{threshold rule}, which prescribes the player to choose $C$ is and only if his private loss $\ell_i$ from cooperating with the defecting player is below a certain threshold.

\subsection{Preliminary Observations}

The next lemma states that in any symmetric equilibrium, an agent's strategy can be fully specified by a \emph{threshold rule} defined by $\ell^{*}(\pi)$. According to this rule, an agent $i$ plays $C$ if and only if his cost $\ell_i$ falls below this threshold:

\begin{lemma}\label{lemma:threshold}
In any symmetric Bayesian Nash equilibrium, an agent's strategy is defined by a function
\begin{equation}\label{eq:threshold}
\ell^{*}:[0,1]\to[0,\overline{\ell}],
\end{equation}
such that\footnote{Given that we assume a continuous distribution for $\ell$, the specific tie-breaking rule at $c=\ell^{*}(\pi)$ is immaterial.}
\begin{equation}\label{eq:threshstrat}
\sigma_i(\ell,\pi)=
\begin{cases}
C&\text{if }\ell\leq \ell^{*}(\pi),\\
D&\text{if }\ell>\ell^{*}(\pi).
\end{cases}
\end{equation}
\end{lemma}

\begin{proof}
See Appendix \ref{proof:threshold}.
\end{proof}

It should be stressed that Lemma \ref{lemma:threshold} suggests that \emph{any} candidate equilibrium strategy is of this type; however, in asymmetric equilibria, the two players could switch their action around two \emph{different} thresholds $\ell_1^{*}$ and $\ell_2^{*}$. In what follows, we will refer to an agent's strategy as simply $\ell^{*}(\pi)$.

\subsection{\label{subs:common}Common beliefs}

First, we consider a situation in which agents share a common belief $\pi\in[0,1]$ about the likelihood of encountering an `honest' type. The results outlined in Proposition \ref{prop:threshold} assume that the distribution function $F(\cdot)$ for losses ($\ell$) satisfies the following regularity condition:
\begin{assumption}[Monotone Hazard Rate]\label{ass:monhr}
Define the hazard rate by
\begin{equation}
h(\ell)\triangleq\frac{f(\ell)}{1-F(\ell)}.
\end{equation}
Assume that $h(\ell)$ is increasing in $\ell$.
\end{assumption}

The equilibrium construction will make use of the function
\begin{equation}\label{eq:psifunction}
\Psi(\ell;\pi)\triangleq\frac{1+m-b}{1-F(\ell)}\cdot\frac{\pi}{1-\pi}-(b-1)\frac{F(\ell)}{1-F(\ell)},
\end{equation}
which can be interpreted as the reduced-form best response---the optimal threshold $\hat{\ell}$, given that player $i$'s opponent uses the threshold $\ell$.

The next proposition establishes the existence of an equilibrium in threshold strategies:

\begin{proposition}
\label{prop:threshold}
Suppose the distribution $F(\cdot)$ satisfies Assumption \ref{ass:monhr}.

In the common beliefs case, the strategy of each agent $\sigma_i(\ell,\pi)$ in any symmetric Bayesian Nash equilibrium can be represented by a function $\ell^{*}_{\text{c}}(\pi)$, so that an agent opts to cooperate if and only if $\ell_i\leq\ell^{*}_{\text{c}}(\pi)$.

If $\overline{\ell}\leq b-1$, the threshold $\ell^{*}_{\text{c}}(\pi)$ is strictly increasing in $\pi$ for $\pi\leq\frac{b-1}{m}$ and is constant at $\overline{\ell}$ for $\pi>\frac{b-1}{m}$.

If $\overline{\ell}>b-1$, the threshold $\ell^{*}_{\text{c}}(\pi)$ is increasing in $\pi$ for $\pi\leq\frac{b-1}{m}$.

There exists a $\pi'>\frac{b-1}{m}$, implicitly defined by
\begin{equation*}
\Psi(\ell';\pi')=\pi',
\end{equation*}
where $\ell'$ satisfies $\ell'-\frac{1}{h(\ell')}=b-1$, such that for $\pi\in\left(\frac{b-1}{m},\pi'\right)$, there exist three equilibria: one where $\ell^{*}_{\text{c}}(\pi)=\overline{\ell}$ and the other two, $\ell_L(\pi)$ and $\ell_H(\pi)$, with $\ell_L(\pi)<\ell_H(\pi)<\overline{\ell}$. Furthermore, the threshold $\ell_L(\pi)$ is increasing in $\pi$ whereas the threshold $\ell_H(\pi)$ is decreasing in $\pi$.
\end{proposition}

\begin{proof}
See Appendix \ref{proof:thresholdprop}.
\end{proof}

This proposition provides an equilibrium characterization for the common beliefs' case. It should be noted that for high $\pi$'s, there is a scope for multiple equilibria. Intuitively, this is because the presence of `honest' types adds an element of a coordination game into the standard Prisoner's dilemma: if both players are strategic, but each attaches a high probability to the event that his partner is honest, both would prefer to play $C$, despite the fact that $D$ is a dominant action in the game when it is commonly known that both players are strategic (that is, where $\pi=0$).

Additionally, let us note that the best response $\hat{\ell}(\pi;\ell)$ is decreasing in $\ell$ for low $\pi$ but starts to increase in $\ell$ for large $\pi$. This means a greater readiness to cooperate by their partner (higher $\ell$) makes the player \emph{less eager} to cooperate (decreases $\hat{\ell}$) when if they hold low $\pi$. However, when the probability to meet an honest partner becomes sufficiently high, the players' cooperation decisions become \emph{complementary} ($\hat{\ell}$ increasing with $\ell$). Notice also that the critical value for $\pi$ above which this transformation (from decreasing to increasing $\hat{\ell}(\pi;\cdot)$) occurs is equal to the benefit-cost ratio from defection, $\frac{b-1}{m}$. If the player were certain to deal with the (cooperating) strategic type, he would get a net gain $(b-1)$ from defection; in case it turned out that his partner is honest, he would suffer a net loss of $b-(b-m)=m$ from defection.

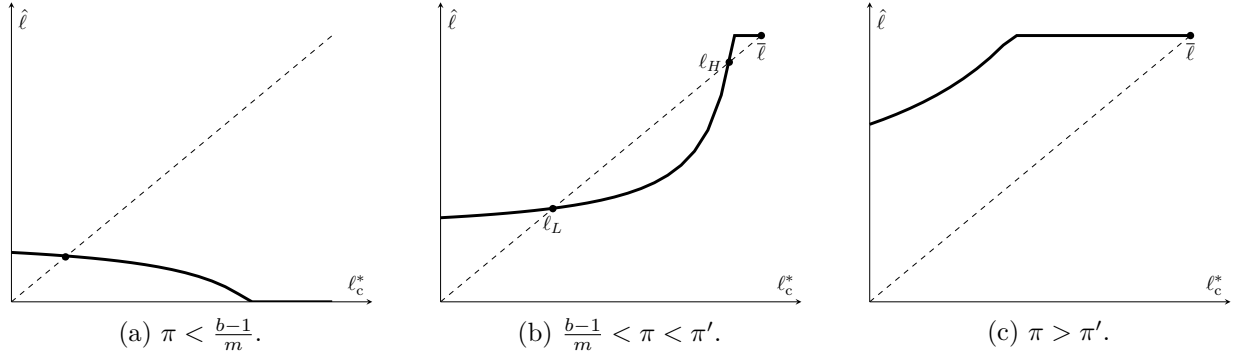
\begin{figure}[t]
\begin{center}
\newcommand\lowpi{%
\begin{tikzpicture}[scale=0.7]
\begin{axis}
                [
                    xmin=0, xmax=9,
                    ymin=0, ymax=9,
                    xlabel={$\ell^{*}_{\text{c}}$},
                    ylabel={$\hat{\ell}$},
                    axis y line=middle,
                    axis x line=middle,
                    xtick=\empty,
                    ytick=\empty,
                    declare function={H(\x,\m,\c,\b,\p)=max(0,min(((1+\m)/(1-\x/\c))*(\p/(1-\p))-(\b-1)*((\x/\c)/(1-\x/\c)),\c));},
                    domain=0:8,no marks
               ]
 \addplot[ultra thick,black]{H(x,47,8,3,0.03)};
 \draw[dashed] (0,0) -- (8,8);
 \fill (1.35,1.35) circle[radius=2pt];
\end{axis}
\end{tikzpicture}
}
\newcommand\middlepi{%
\begin{tikzpicture}[scale=0.7]
\begin{axis}
                [
                    xmin=0, xmax=9,
                    ymin=0, ymax=9,
                    xlabel={$\ell^{*}_{\text{c}}$},
                    ylabel={$\hat{\ell}$},
                    axis y line=middle,
                    axis x line=middle,
                    xtick=\empty,
                    ytick=\empty,
                    declare function={H(\x,\m,\c,\b,\p)=max(0,min(((1+\m)/(1-\x/\c))*(\p/(1-\p))-(\b-1)*((\x/\c)/(1-\x/\c)),\c));},
                    domain=0:8,no marks
               ]
 \addplot[ultra thick,black]{H(x,47,8,3,0.05)};
 \draw[dashed] (0,0) -- (8,8);
 \fill (2.8,2.8) circle[radius=2pt] node[below]{$\ell_L$};
  \fill (7.2,7.2) circle[radius=2pt] node[left]{$\ell_H$};
   \fill (8,8) circle[radius=2pt] node[below]{$\overline{\ell}$};
\end{axis}
\end{tikzpicture}
}
\newcommand\highpi{%
\begin{tikzpicture}[scale=0.7]
\begin{axis}
                [
                    xmin=0, xmax=9,
                    ymin=0, ymax=9,
                    xlabel={$\ell^{*}_{\text{c}}$},
                    ylabel={$\hat{\ell}$},
                    axis y line=middle,
                    axis x line=middle,
                    xtick=\empty,
                    ytick=\empty,
                    declare function={H(\x,\m,\c,\b,\p)=max(0,min(((1+\m)/(1-\x/\c))*(\p/(1-\p))-(\b-1)*((\x/\c)/(1-\x/\c)),\c));},
                    domain=0:8,no marks
               ]
 \addplot[ultra thick,black]{H(x,47,8,3,0.1)};
 \draw[dashed] (0,0) -- (8,8);
   \fill (8,8) circle[radius=2pt] node[below]{$\overline{\ell}$};
\end{axis}
\end{tikzpicture}
}
\subfloat[$\pi<\frac{b-1}{m}$.]{\lowpi}
\qquad
\subfloat[$\frac{b-1}{m}<\pi<\pi'$.]{\middlepi}
\qquad
\subfloat[$\pi>\pi'$.]{\highpi}
\caption{Player $i$'s best response and the determination of equilibrium $\hat{\ell}(\pi;\ell^{*}(\pi))=\ell^{*}_{\text{c}}(\pi)$ ($m=50$, $b=3$, $\ell\sim U[0,\overline{\ell}]$ with $\overline{\ell}=8$).}
\label{fig:eqthresh}
\end{center}
\end{figure}

Figure \ref{fig:eqthresh} depicts the equilibrium thresholds for different ranges of $\pi$:

\begin{itemize}
\item \textbf{Panel (a)} corresponds to $\pi<\frac{b-1}{m}$. Here, the intersection between the agent's best response (solid curve) and the line $\ell^{*} = \hat{\ell}$ in unique.

\item \textbf{Panel (b)} corresponds to $\frac{b-1}{m}<\pi<\pi'$. In this case, there exist three symmetric equilibria at $\ell_L$, $\ell_H$ and $\overline{\ell}$.

\item \textbf{Panel (c)} corresponds to $\pi>\pi'$. In this case, the only intersection between the best response curve and the 45-degree line occurs at $\hat{\ell}=\overline{\ell}$, suggesting that players cooperate for any $\ell$.
\end{itemize}

\subsection{\label{subs:diverse}Diverse beliefs}

When an agent faces uncertainty regarding his partner's belief $\pi_j$, the players can no longer coordinate on $\pi$. In such circumstances, it would be more convenient to represent the agent's strategy by $\pi^{*}_{\text{d}}(\ell)$, the threshold value for the belief $\pi_i$, where agent $i$ engages in cooperation if and only if $\pi_i\geq\pi^{*}_{\text{d}}(\ell)$. Since the net benefit from cooperating is strictly increasing in the probability of facing an honest partner, for any $\ell$, there exists at most one $\pi^{*}$ at which the player is indifferent between $C$ and $D$.

The following lemma outlines the behaviour of $\pi^{*}_{\text{d}}(\ell)$.

\begin{lemma}
\label{lem:divthresh}
In any symmetric equilibrium, the agent's strategy can be described by
\begin{equation}
\label{eq:divthres}
\sigma_i(\ell,\pi)=
\begin{cases}
C&\text{if }\pi\geq\pi^{*}_{\text{d}}(\ell),\\
D&\text{if }\pi<\pi^{*}_{\text{d}}(\ell).
\end{cases}
\end{equation}
\end{lemma}

\begin{proof}
See Appendix \ref{proof:lemdivthresh}.
\end{proof}

Essentially, Lemma \ref{lem:divthresh} suggests that the entire $(\pi,\ell)$ space can be partitioned into the two regions separated by the curve $\pi^{*}_{\text{d}}(\ell)$. If the player's own $(\pi,\ell)$ falls within the upper-left region, he chooses to defect, while in the lower-right region he will prefer to cooperate. One would expect that the function $\pi^{*}_{\text{d}}(\ell)$ is increasing, implying that a higher cost of cooperating ($\ell$) with the player who defects necessitates a higher level of faith in the counterpart's honesty ($\pi$) for an agent to be willing to cooperate.

This is established in the following proposition:
\begin{proposition}
\label{main:diverse}
There exists a unique $\pi^{*}_{\text{d}}(\ell)$, implicitly defined by
\begin{equation}
\label{eq:propthresh}
\pi^{*}_{\text{d}}(\ell)=1-\frac{1+m-b}{m+(\ell-(b-1))\int_0^{\overline{\ell}}G(\pi^{*}_{\text{d}}(\ell))\dd F(\ell)},
\end{equation}
which is strictly increasing in $\ell$.
\end{proposition}

\begin{proof}
See Appendix \ref{proof:maindiv}.
\end{proof}

In the proof of Proposition \ref{main:diverse}, we take an \emph{arbitrary} function $\pi(\ell)$ and iteratively applying the operator given by the right-hand side of \eqref{eq:propthresh}, showing that there exists a unique fixed point.

\section{\label{sec:cooperation}Determining the Probability of Cooperation}

Having characterized the symmetric equilibria in the common and diverse beliefs' cases, we can now address the central question of our paper: How does belief diversity affect the likelihood of a particular player choosing to cooperate? We address this question from an \emph{ex post} and an \emph{ex ante} perspective.

First, we compute the probability that the player will cooperate, \emph{conditional on his belief $\pi$} (ex post perspective). Imagine that someone observes the society where $\pi$ is known---in that context, $\pi$ might be interpreted as the actual fraction of honest types in the population. Conditional on $\pi$, the probability that the player cooperates is equal to the probability that his cost is below the threshold, that is, $\ell_i\leq\ell^{*}(\pi)$, which happens with probability $F(\ell^{*}(\pi))$. In that situation, the \emph{common beliefs'} scenario corresponds to the situation when both players know $\pi$ and play according to $\ell^{*}_{\text{c}}(\pi)$. By contrast, the \emph{diverse beliefs'} scenario corresponds to the situation when both players still hold beliefs $\pi_1=\pi_2=\pi$, but \emph{they do not know that}. Since their strategies are described by the threshold $\pi^{*}_{\text{d}}(\ell)$ as given by \eqref{eq:propthresh}, and since this function is strictly increasing in $\ell$, we can invert that and then apply the $F(\cdot)$ function to compute the likelihood of cooperation.

Second, we compute the probability of cooperation from the viewpoint of someone who \emph{does not observe $\pi$} (ex ante perspective). In that case, the knowledge of the $\ell^{*}_{\text{c}}(\pi)$ function for the common beliefs' case and of the $\pi^{*}_{\text{d}}(\ell)$ for the diverse beliefs' case allows us to compute the probability of cooperation as $\int_0^1F(\ell^{*}_{\text{c}}(\pi))\dd G(\pi)$ for the former case and as $\int_0^{\overline{\ell}}[1-G(\pi^{*}_{\text{d}}(\ell))]\dd F(\ell)$ for the latter.

For illustrative purposes, we focus on a particular setting where both the costs $\ell$ and the agents' beliefs $\pi$ are uniformly distributed on $[0,1]$. Furthermore, we restrict our analysis to the case $b\geq2=\overline{\ell}+1$, ensuring a unique equilibrium in the common beliefs' scenario (as stated in Proposition \ref{prop:threshold}).

\subsection{Probability of cooperation from an ex post perspective}

In this section, we assume that the belief $\pi$ is known to the outside observer (but not necessarily to his partner). From that perspective, we will compute the probability of cooperation in the common and diverse beliefs' settings, for a given $\pi$, and compare them. The following proposition provides the equilibrium thresholds within this specific parametric context:

\begin{proposition}
\label{prop:heterogeneous}

Consider the parametric case with $\ell_j\sim U[0,1]$ and $\pi_j\sim U[0,1]$ and assume that $b\geq2$.

\begin{enumerate}
\item In a common beliefs' setting, the unique threshold can be expressed as:
\begin{equation}
\label{eq:combackthr}
\ell^{*}_{\text{c}}(\pi)=
\begin{dcases}
\frac{b}{2}-\sqrt{\frac{b^2}{4}-(1+m-b)\frac{\pi}{1-\pi}},&\text{if }\pi<\frac{b-1}{m},\\
1&\text{if }\pi\geq\frac{b-1}{m}.
\end{dcases}
\end{equation}
\item If $m\gg b$, the threshold for a diverse beliefs' setting can be approximated as:
\begin{equation}
\label{eq:divbackthr}
\ell^{*}_{\text{d}}(\pi)=
\begin{dcases}
0&\text{if }\pi<\beta,\\
\frac{1}{\beta}\left(\frac{1+m-b}{1-\pi}-\alpha\right),&\text{if }\beta\leq\pi<1-\frac{1+m-b}{\alpha+\beta},\\
1&\text{if }\pi\geq1-\frac{1+m-b}{\alpha+\beta},
\end{dcases}
\end{equation}
where $\alpha$ and $\beta$ are defined as follows:
\begin{equation}
\label{eq:alphabeta}
\begin{split}
\alpha&=\frac{1+m-b}{2}\left(1+\sqrt{1+\frac{4(b-1)}{1+m-b}}\right)\\
\beta&=\frac{\sqrt{1+\frac{4(b-1)}{1+m-b}}-1}{\sqrt{1+\frac{4(b-1)}{1+m-b}}+1}
\end{split}
\end{equation}
\end{enumerate}
\end{proposition}

\begin{proof}
See Appendix \ref{proof:heterogen}.
\end{proof}

\begin{figure}[t]
\begin{center}
\newcommand\com{%
\begin{tikzpicture}[scale=1]
\begin{axis}
                [
                    xmin=0, xmax=1.2,
                    ymin=0, ymax=1.2,
                    xlabel={$\pi$},
                    ylabel={$\ell^{*}$},
                    xtick={1},
                    ytick={1},
                    xlabel style={at={(axis description cs:0.5,-0.1)},anchor=north},
                    ylabel style={at={(axis description cs:-0.1,0.5)},rotate=90,anchor=south},
                    axis y line=middle,
                    axis x line=middle,
                    declare function={C(\x,\m,\b)=max(0,min(\b/2-(\b^2/4-(1+\m)*(\x/(1-\x)))^(1/2),1));},
                    domain=0:0.99,no marks
               ]
 \addplot[ultra thick,black]{C(x,7,18)};
 \draw[dotted] (0,1) node[left]{1} -- (1,1) -- (1,0) node[below]{1};
\end{axis}
\end{tikzpicture}
}

\newcommand\het{%
\begin{tikzpicture}[scale=1]
\begin{axis}
                [
                    xmin=0, xmax=1.2,
                    ymin=0, ymax=1.2,
                    xtick={1},
                    ytick={1},
                    xlabel={$\pi$},
                    ylabel={$\ell^{*}$},
                    xlabel style={at={(axis description cs:0.5,-0.1)},anchor=north},
                    ylabel style={at={(axis description cs:-0.1,0.5)},rotate=90,anchor=south},
                    axis y line=middle,
                    axis x line=middle,
                    declare function={H(\x,\a,\b,\m)=max(0,min(((1+\m)/(1-\x)-\a)*1/\b,1));},
                    domain=0:0.99,no marks
               ]
 \addplot[ultra thick,black]{H(x,16.2,0.5,7)};
 \draw[dotted] (0,1) node[left]{1} -- (1,1) -- (1,0) node[below]{1};
\end{axis}
\end{tikzpicture}
}

\subfloat[Common beliefs.]{\com}
\qquad
\subfloat[Diverse beliefs.]{\het}

\caption{Equilibrium thresholds for the common and diverse beliefs.}
\label{fig:twothresh}
\end{center}
\end{figure}
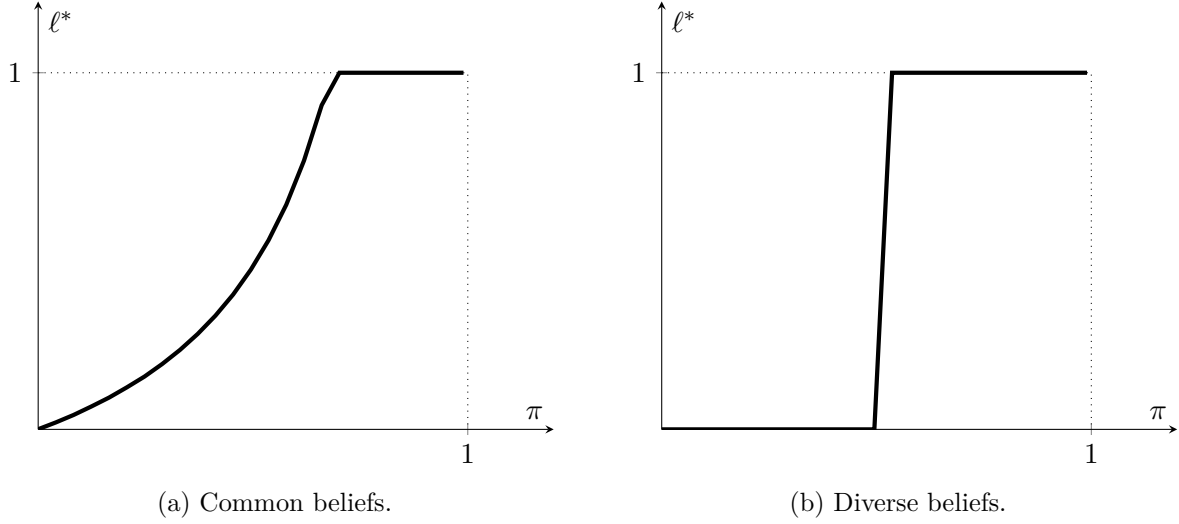

For the diverse beliefs' case, it is easier to work with the inverted function, $\ell_{\text{d}}^{*}(\pi)$, which shows the threshold for the costs, below which the player switches to cooperation. Comparison of the two thresholds, $\ell_{\text{c}}^{*}(\pi)$ and $\ell_{\text{d}}^{*}(\pi)$, allows us to directly compare the probabilities of cooperation.\footnote{Recall that $F(\ell)=\ell$, since $\ell\sim U[0,1]$.} There are two things to notice. First, $\ell_{\text{c}}^{*}(\pi)$ is strictly positive while $\ell_{\text{d}}^{*}(\pi)$ remains equal to zero for very low values of $\pi$. This suggests that cooperation is possible even at the very low values of $\pi$'s, provided that the belief to meet an honest partner is a common knowledge. This allows the players can \emph{coordinate} their behavior on $\pi$, facilitating cooperation. This coordination is no longer possible if player $i$ does not observe his partner's belief.\footnote{Note that in the diverse beliefs' case, we assume $\pi_1\indep\pi_2$, so that my own belief that my partner is honest does not tell me anything about my partner's belief that I am honest.}

However, for $\pi>\beta$, the threshold for the diverse beliefs' case, $\ell_{\text{d}}^{*}(\pi)$, starts to increase, and does so very quickly. It can be shown that it reaches the upper bound of one \emph{before} $\ell_{\text{c}}^{*}(\pi)$ does.\footnote{Straightforward computations show that $1-\frac{1+m-b}{\alpha+\beta}<\frac{b-1}{m}$; for details, see the \hyperref[proof:dagger]{proof} of Proposition \ref{prop:dagger}.} This suggests that the possibility to coordinate on $\pi$ becomes both a blessing and a curse: if I have a low trust in my partner, that is, if my belief $\pi$ that he is honest is low, then knowing that \emph{he thinks exactly as I do}---that is, we are in the common beliefs' scenario---induces me to cooperate more: $\ell_{\text{c}}^{*}(\pi)>\ell_{\text{d}}^{*}(\pi)$. For the same reason, however, when $\pi$ is relatively high (close to, but still less than $\frac{b-1}{m}$), the common knowledge of $\pi$ \emph{impedes} cooperation: introducing belief uncertainty about $\pi$ increases the probability of cooperation by making players more difficult to coordinate on defection.

We may juxtapose the two Figures \ref{fig:twothresh}(a) and \ref{fig:twothresh}(b) to see that for $\pi<\frac{b-1}{m}$, these curves intersect exactly once.\footnote{For $\pi\geq\frac{b-1}{m}$, the two curves coincide at $\ell_{\text{c}}^{*}(\pi)=\ell_{\text{d}}^{*}(\pi)=1$.} This implies that there exists a unique belief $\pi=\pi^{\dagger}$ such that cooperation becomes more likely in the diverse beliefs' case. The next proposition formally characterizes this critical value for $\pi$.

\begin{proposition}
\label{prop:dagger}
There exists a unique value, denoted as $\pi^{\dagger}$, such that for all $\pi>(<)\pi^{\dagger}$, the probability of cooperation is higher (lower) in the context of diverse backgrounds when compared to the common background setting.
\end{proposition}

\begin{proof}
See Appendix \ref{proof:dagger}.
\end{proof}

\begin{figure}[t]
\newcommand\pib{%
\includegraphics[scale=0.25]{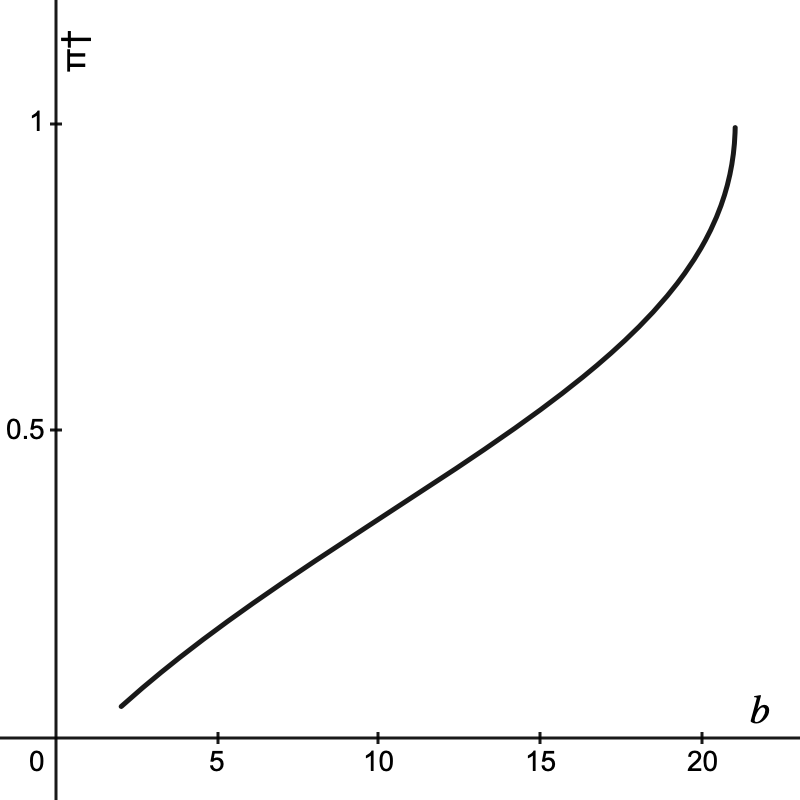}
}
\newcommand\pim{%
\includegraphics[scale=0.25]{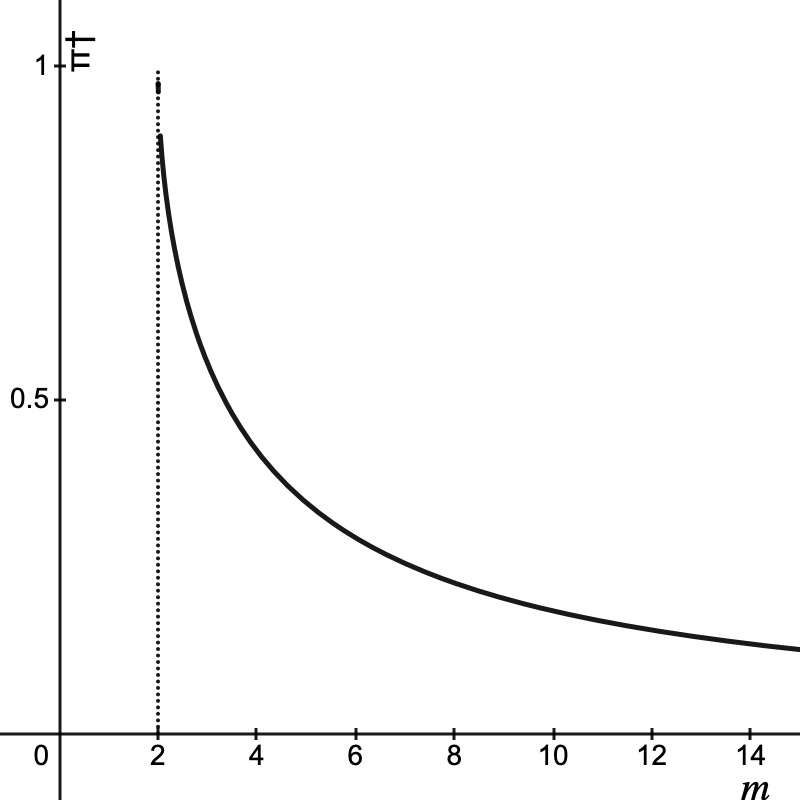}
}
\subfloat[$\pi^{\dagger}$ is increasing in $b$ ($m=20$).]{\pib}
\qquad
\subfloat[$\pi^{\dagger}$ is decreasing in $m$ ($b=3$).]{\pim}
\caption{The critical value $\pi^{\dagger}$ is increasing in $b$ and decreasing in $m$.}
\label{fig:pimg}
\end{figure}

The expression \ref{eq:pidagimpl} in the proof implicitly determines $\pi^{\dagger}=\pi^{\dagger}(b,m)$ as an implicit function of the model primitives, the (gross) benefit to defecting a strategic partner ($b$) and the (gross) cost of defecting an honest partner ($m$). The Implicit Function Theorem can then be used to analyze how $\pi^{\dagger}$ changes with $b$ and $m$. While it is quite tedious to compute exact expressions for $\frac{\dd\pi^{\dagger}}{\dd b}$ and $\frac{\dd\pi^{\dagger}}{\dd m}$, visual analysis suggests that $\frac{\dd\pi^{\dagger}}{\dd b}>0$ and $\frac{\dd\pi^{\dagger}}{\dd m}<0$. Figure \ref{fig:pimg} illustrates the impact of $b$ on $\pi^{\dagger}$.

Since by definition of $\pi^{\dagger}$, the range of beliefs for which common knowledge of $\pi$ increases the likelihood of cooperation is given by $(0,\pi^{\dagger})$, the decrease (increase) in $\pi^{\dagger}$ can be interpreted as greater (lower) benefits to belief diversity---in terms of increasing the scope for cooperation. Put differently, one would expect that the common knowledge of $\pi$ will foster cooperation in environments where the gains from defecting are large (high $b$) and the moral costs are negligible (low $m-b$).

\subsection{Probability of cooperation from an ex ante perspective}

So far, we have considered the impact of belief diversity on cooperation \emph{for a given $\pi$}. The conclusion reached by Proposition \ref{prop:dagger} suggests that for $\pi>\pi^{\dagger}$, belief diversity raises the scope for cooperation, whereas for $\pi<\pi^{\dagger}$, cooperation is more likely in the environment where $\pi$ is commonly known.

This raises the question whether belief diversity raises or lowers the scope for cooperation from an \emph{ex ante} perspective---that is, from the viewpoint of an outside observer who does know $\pi$. To that end, we consider the ex-ante probability of cooperation in the environment $k\in\{\text{common},\text{diverse}\}$ as
\begin{equation}
p_k=\int_{0}^{\overline{\ell}}[1-G(\pi^{*}_k(\ell))]\mathrm{d}F(\ell).
\end{equation}
We would like to know how the model primitives, the benefit to defection $b$ and the cost of mistreating the `honest' partner $m$ affect the benefits from belief diversity.

Our next proposition characterizes these two probabilities in the context of our parametric example with the uniformly distributed costs $\ell$ and beliefs $\pi$.

\begin{proposition}
\label{prop:exante}
Consider the case where $\ell,\pi\sim U[0,1]$. Define
\begin{equation}
\varphi\equiv\sqrt{1+m-b+\frac{b^2}{4}} \quad \text{and} \quad \gamma\equiv\sqrt{1+\frac{4(b-1)}{1+m-b}}.
\end{equation}

The \emph{ex ante} probabilities of cooperation for $k\in\{\text{common},\text{diverse}\}$ are given by
\begin{flalign}
\label{eq:forpc}
p_{\text{c}}&=\frac{1+m-b}{2\varphi}\ln\left(1+\frac{2\varphi}{\varphi(\varphi-1)-\frac{b}{2}\left(\frac{b}{2}-1\right)}\right)\\
\label{eq:forpd}
p_{\text{d}}&=(1+m-b)\frac{\gamma+1}{\gamma-1}\ln\left(1+\frac{2(\gamma-1)}{(1+m-b)(\gamma+1)^2}\right)
\end{flalign}
\end{proposition}

\begin{proof}
See Appendix \ref{proof:exante}.
\end{proof}

\begin{figure}[t]
\includegraphics[scale=0.5]{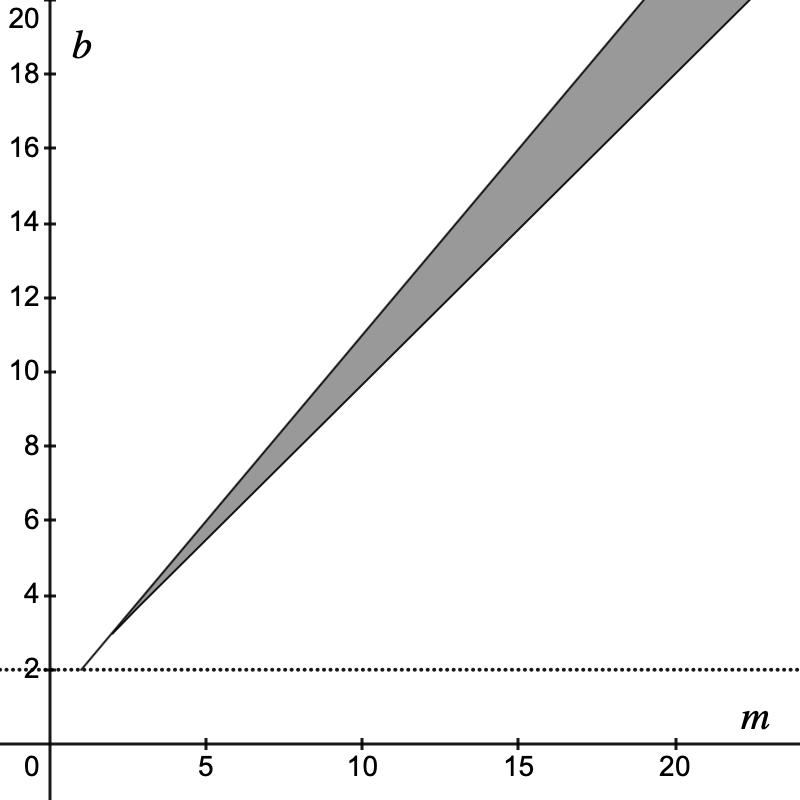}
\caption{The shaded area delineates the region where $p_{\text{d}}>p_{\text{c}}$.}
\label{fig:exante}
\end{figure}

The two expressions for $p_{\text{c}}$ and $p_{\text{d}}$ allow us to represent graphically the region in the $(m,b)$ space where the probability of cooperation is more likely in an environment with belief diversity, that is, $p_{\text{d}}>p_{\text{c}}$. Figure \ref{fig:exante} represents this region. Keep in mind that Assumption \ref{ass:par} implies that we restrict ourselves to a subset of $m$ and $b$ such that $b<m+1$, which corresponds to the steeper line at the left border of the shaded region.

The graph suggests that, for each $b$, there exists an open interval $M(b)=(\underline{m}(b),\overline{m}(b))$, such that belief diversity increases the probability of cooperation if and only if $m\in M(b)$. Notice that, for extremely high moral costs $m$, $\pi^{\dagger}$ approaches very close to $\frac{b-1}{m}$, the value of $\pi$ above which we have both $\ell_{\text{c}}^{*}(\pi)=\ell_{\text{d}}^{*}(\pi)=1$, which means that for almost all $\pi$'s, we have $\ell_{\text{c}}^{*}(\pi)>\ell_{\text{d}}^{*}(\pi)=1$, and thus $p_{\text{c}}>p_{\text{d}}$. Intuitively, the benefits from coordinating on $\pi$ in the setting where $\pi$ is commonly known outweighs the benefits from belief diversity whenever the moral cost $m$ is very high.

\section{\label{sec:applications}Discussion: Extensions and Directions for Future Work}

In this section, we outline several ways to extend our setup and discuss potential experimental tests of our paper's implications.

First, we explore the implications of our setup for the case where two players have heterogeneous beliefs, $\pi_1$ and $\pi_2$. We demonstrate that if one player has a sufficiently low belief in their partner's honesty, an increase in their \emph{partner's} belief may lead that player to cooperate \emph{less} frequently (Proposition \ref{prop:hetbel}). This indicates that greater optimism can backfire in an environment with commonly known beliefs, where a player's increased expectation of encountering an honest partner enhances the strategic partner's opportunism.

Second, we consider scenarios involving more than two agents. Specifically, we examine the behavior of one strategic player in a group consisting of $n$ other players, each potentially honest. In this context, choosing to defect can be likened to violating a taboo---committing an act unthinkable to the group members.\footnote{This aligns with our definition of `honesty' as the incapacity to commit the socially detrimental action $D$.} This exploration allows us to assess how group size influences the benefits of belief diversity.

Third, we plan to adapt our model to include the factor of reputation, effectively endogenizing the moral cost parameter $m$. In this revised model, we introduce a multi-period setting where an agent's past decisions are recorded, making reputation a pivotal aspect of their strategic persona.

We conclude this section by proposing methods for experimentally testing our model's implications.

\subsection{Heterogeneous Beliefs}

In some cases, it may be appropriate to consider two players with \emph{different} beliefs, both aware of this difference.\footnote{We are grateful to a referee for suggesting we explore this scenario.} For instance, in interactions between a military combatant and a civilian, the former may be more skeptical about the latter's intentions. Interestingly, the equilibrium analysis in Section \ref{subs:common} indicates that an increase in one player's belief in their partner's honesty may decrease the partner's inclination to cooperate.

To illustrate, let's briefly review a scenario where two players hold distinct beliefs $\pi_1$ and $\pi_2>\pi_1$, with both players fully aware of these beliefs. Although different beliefs disrupt symmetry, each player's strategy still follows a threshold pattern, as indicated by Lemma \ref{lemma:threshold}.\footnote{While Lemma \ref{lemma:threshold} is formulated for \emph{symmetric} Bayesian Nash equilibrium, our \hyperref[proof:threshold]{proof} clarifies that in \emph{any} equilibrium, player $i$'s optimal response always adheres to a cutoff rule for $\ell$.}

As Proposition \ref{prop:threshold} and the analysis of Figure \ref{fig:eqthresh} suggest, a player's optimal response $\hat{\ell}(\pi;\cdot)$ \emph{decreases} with their partner's threshold for low $\pi$ values and \emph{increases} for high $\pi$ values. Thus, if $\pi_1<\frac{b-1}{m}$ and $\pi_2>\frac{b-1}{m}$, an increase in $\pi_2$ will extend the `optimist' (player 2 with $\pi_2$)'s best response outward, consequently diminishing the `pessimist' (player 1 with belief $\pi_1$)'s propensity to cooperate.

This is summarised in the following proposition:
\begin{proposition}
\label{prop:hetbel}
Consider a setting in which the players hold asymmetric, commonly known beliefs $\pi_1$ and $\pi_2$. If $\pi_1<\frac{b-1}{m}<\pi_2$, then an increase in $\pi_2$ will reduce the equilibrium threshold $\hat{\ell}_1(\pi_1)$ and increase the equilibrium threshold $\hat{\ell}_2(\pi_2)$.
\end{proposition}

\begin{proof}
See Appendix \ref{proof:hetbel}.
\end{proof}

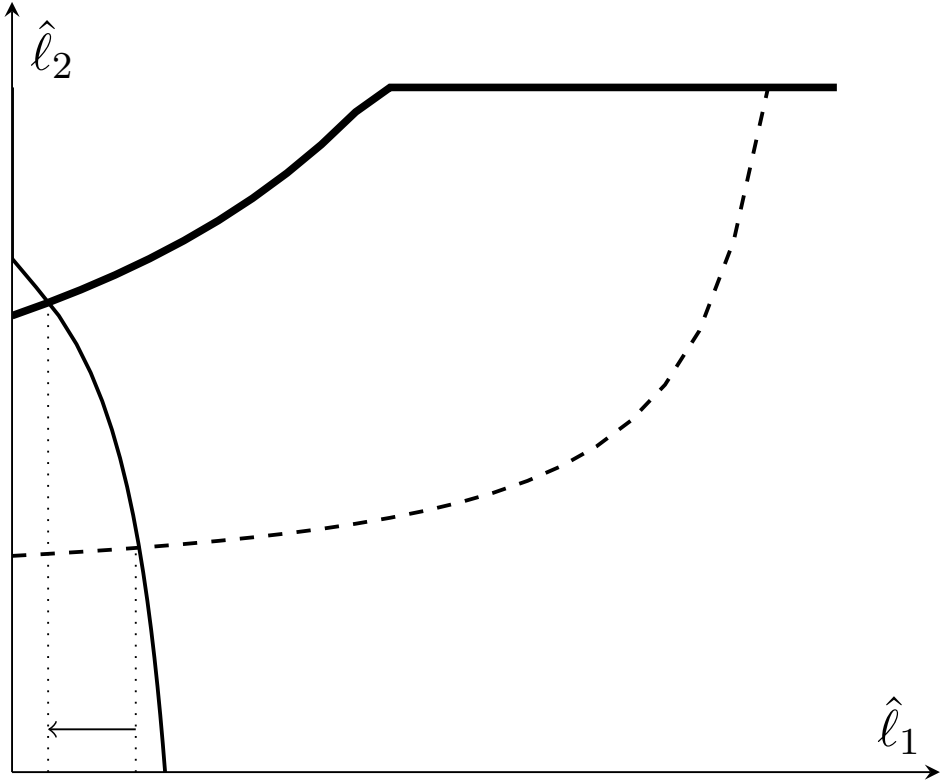
\begin{figure}
\begin{tikzpicture}[scale=1.8]
\begin{axis}
                [
                    xmin=0, xmax=9,
                    ymin=0, ymax=9,
                    xlabel={$\hat{\ell}_1$},
                    ylabel={$\hat{\ell}_2$},
                    axis y line=middle,
                    axis x line=middle,
                    xtick=\empty,
                    ytick=\empty,
                    declare function={H(\t,\m,\c,\b,\p)=max(0,min(((1+\m)/(1-\t/\c))*(\p/(1-\p))-(\b-1)*((\t/\c)/(1-\t/\c)),\c));},
                    domain=0:8,no marks
                ]
    \addplot[thick,black,variable=\t]({H(t,47,8,3,0.03)},{t});

    \addplot[thick,black,dashed]({x},{H(x,47,8,3,0.05)});

    \addplot[ultra thick,black]({x},{H(x,47,8,3,0.1)});
    \draw[dotted] (0.35,0) -- (0.35,5.5);
    \draw[dotted] (1.2,0) -- (1.2,2.6);
    \draw[->] (1.2,0.5) -- (0.35,0.5);
\end{axis}
\end{tikzpicture}
\caption{The increase in $\pi_2$ reduces $\hat{\ell}_1$.}
\label{fig:asymm}
\end{figure}

The idea behind Proposition \ref{prop:hetbel} is demonstrated in Figure \ref{fig:asymm}. The initial equilibrium is represented by an intersection of the decreasing solid curve for $\hat{\ell}_1(\cdot)$ and the increasing dashed curve for $\hat{\ell}_2(\cdot)$. The increase in $\pi_2$ results in an upward shift of the reaction curve for player 2 (the thick solid increasing curve), and we see that the new equilibrium corresponds to an intersection point above and to the left, that is, $\hat{\ell}_1^{*}$ gets reduced.

Intuitively, this occurs because the `pessimistic' player 1 takes his partner's greater willingness to cooperate as an invitation to defect: for player 1, eagerness cooperation decisions are seen as strategic substitutes. Hence, the increase in player 2's trust that his partner is honest induces the (strategic type) player 1 to cooperate less frequently.

\subsection{Cooperation in Large Groups}

So far our focus has remained squarely on bilateral interactions in which randomly paired agents choose between cooperation and defection. However, as long as we interpret $m$ as the moral cost coming from the perception of a certain type of behavior as `completely unacceptable,' it would be instructive to think of how our setup might be used to analyze the maintenance of certain social practices, whose violation might be regarded as a taboo.\footnote{\cite{JindaniYoung2020} provide an account of the dynamics of social norms in an environment where costly practices may be reproduced due to the preference for conformity.}

In this section, we consider the defection choice of a single individual who might contemplate the possibility of being in a group consisting entirely of honest people. In that circumstance, defection might be regarded as a serious offense, a violation of a taboo, leading to an ostracism. We are interested in how belief diversity impacts incentives to cooperate in large groups.

Consider a meeting of $n+1$ agents, each deciding whether to cooperate or defect. If all agents opt for cooperation, each participant receives a payoff of $1$. Conversely, if at least one agent defects, any cooperating agent will incur their cost $\ell$, while each defecting agent receives a payoff of $b$ -- provided at least one other agent in the interaction is also strategic. However, if all other agents in the group are 'honest,' the defector must bear a moral cost $m$.

In the common beliefs' setting, the agent's strategy will be given by the threshold implicitly defined by
\begin{equation}
\label{eq:multc}
(1-\ell^n{\text{c}}(\pi))\sum_{k=0}^{n}\binom{n}{k}\pi^{k}[(1-\pi)F(\ell^n{\text{c}}(\pi))]^{n-k}-\ell^n{\text{c}}(\pi)=b-m\pi^n.
\end{equation}

In the diverse beliefs' scenario, the agent's decision will be implicitly defined by
\begin{equation}
\label{eq:multd}
(1-\ell^n_{\text{d}}(\pi))\sum_{k=0}^{n}\binom{n}{k}\pi^k\left[(1-\pi)\int_0^1F(\ell_{\text{d}}^n(\pi))\dd G(\pi)\right]^{n-k}-\ell^n{\text{d}}(\pi)=b-m\pi^n
\end{equation}

In both expressions \eqref{eq:multc} and \eqref{eq:multd}, the left-hand side represents the expected benefit of cooperation by an opportunistic agent who holds belief $\pi$, while the right-hand side stands for his expected benefit from defection. In both cases, the latter is just equal to the (gross) benefit from defection $b$ minus the moral cost $m$ times the probability that \emph{all} other agents in a group are honest, $\pi^n$. The only difference between the two expressions comes from the term that shows up under the summation sign: under the common beliefs' setting, the probability that each particular strategic player cooperates is given by $F(\ell^n{\text{c}}(\pi))$, whereas in the diverse beliefs' setting, it is equal to $\int_0^1F(\ell_{\text{d}}^n(\pi))\dd G(\pi)$. A key point of interest in this setting would be to examine the rate at which the probability of cooperation changes with the increase $n$.

\subsection{\label{subs:repext}Endogenizing $m$ through Reputation Dynamics}

The origins of the moral cost associated with mistreating an honest partner are worth considering. For instance, in the case of the military committing war crimes, it is reasonable to expect that retribution might occur in the future if acts of violence against civilians become public, potentially severely tarnishing their reputation. In this section, we explore how moral costs can arise organically through reputation dynamics.

Imagine a multi-period setting indexed by time $t=0,1,2,\ldots$. An agent enters this environment at time $t=0$ with an initial reputation $\pi$, reflecting the public belief in their honesty. At each time step, the agent is randomly matched with a partner whose tenure is denoted by $\tau\in{0,1,\ldots,t}$, indicating that this partner entered the setting at time $t-\tau$. The partner either maintains an unblemished record---having never defected against an honest agent---or is identified as strategic, thereby diminishing their reputation.\footnote{For simplicity, our model assumes that agents are unaware of past defections against their partner. Unless they can observe the actual losses, $\ell_i$, from agent $i$ in previous interactions, such information is considered irrelevant.}

Each agent has a fixed survival probability $\delta\in[0,1)$ per period, which also serves as the discount factor. An agent at age $t$ with a clean record will have a reputation $\pi_t$. Let's assume this agent is randomly paired with an opponent of age $\tau$, who also has an unblemished record.\footnote{The optimal strategy when facing an agent with a tarnished reputation is to choose $D$, yielding a payoff of $0$.}

The player's optimal decision can be recursively defined as follows:
\begin{equation}
\begin{split}
V(\pi_t;\ell)=\max\bigg\{&\pi_{\tau}+(1-\pi_{\tau})p(\pi_{\tau})-\ell(1-\pi_{\tau})(1-p(\pi_{\tau}))+\delta V(\pi_{t+1};\ell),\\
&(1-\pi_{\tau})p(\pi_{\tau})b+\delta(1-\pi_{\tau})V(\pi_{t+1};\ell)\bigg\},
\end{split}
\end{equation}
where $p(\pi_{\tau})$ signifies the probability that a partner of age $\tau$ will cooperate with the player.

In this setup, the agent incurs no explicit moral cost. However, if the defector's behavior becomes public knowledge, his future payoff becomes zero, as no one would wish to engage with him subsequently. This concept relates to \cite{HirshleiferRasmusen1989}, which offers an insightful perspective on promoting cooperation in groups through the mechanism of ostracism.

\subsection{\label{sec:experdes}Implications for Experimental Study}

In this section, we outline how the implications of our model could be tested in a laboratory setting. Participants could be randomly paired to play the game, given specific values of $b$, $m$, and $\ell$.\footnote{In our model, $\ell$ is considered private information to the player, a simplification to avoid dealing with mixed strategies and to allow for a clearer equilibrium characterization. We posit that in experimental contexts, the inherent heterogeneities among participants, both observable and unobservable, render the choice between playing $C$ and $D$ inconsequential.} Participants would be informed that there is a certain probability that they will encounter a fictitious player (e.g., a computer algorithm) that always cooperates, and defecting against this player results in a loss ($m$).

For the common beliefs scenario, the experimenter would disclose this probability ($\pi$) to the participants beforehand. In the case of diverse beliefs, the experimenter might simply state that there is "some" probability each player could be paired with a fictitious, honest partner. Additionally, in this latter scenario, the experimenter could attempt to infer the participants' beliefs by asking those who chose to cooperate their perceived likelihood of their partner being honest.

A straightforward hypothesis to test, as suggested by Proposition \ref{prop:dagger}, is that relative to the diverse beliefs scenario, publicly revealing a $\pi$ value is likely to increase cooperation in situations where it was initially low and decrease the potential for cooperation in situations where it was initially high.

\section{\label{sec:conclusion}Conclusion}

This paper has explored the likelihood of cooperation in environments where some individuals are inherently honest---consistently cooperating without self-interest---and the potential repercussions for those exploiting this trust.

We have analyzed two scenarios: one where players share a belief about interacting with an honest partner, and another where their beliefs are diverse. The resulting trade-offs indicate that at sufficiently high levels of trust, cooperation incentives become complementary, shifting the dynamic from a Prisoner's Dilemma to a coordination game. Our findings demonstrate that common knowledge of belief promotes coordination on cooperation when beliefs are low, but similarly facilitates coordination on mutual defection when beliefs are relatively high.

Although we treated the moral cost as an exogenous factor, we proposed several ways it could be endogenized. For instance, players may face a reputational cost in multi-period interactions, or the cost could derive from guilt aversion or the fear of disappointing others, particularly when interactions are preceded by a phase of communication.

While our analysis did not extend to the normative implications of our model, our results suggest that disseminating information about the presence of honest individuals in a population could foster cooperative behavior.

\begin{appendices}

\section{Proofs}

\subsection{\label{proof:threshold} Proof of Lemma \ref{lemma:threshold}}
\begin{proof}
The expected payoff for player $i$ when they cooperate, taking into account their own strategy, the strategy of the opponent, and their own type, can be formulated as
\begin{equation}
\label{eq:coop}
u_i(C,\sigma';\ell,\pi)=\pi+(1-\pi)p(\sigma')-(1-\pi)(1-p(\sigma'))\ell,
\end{equation}
while their expected payoff from defection is
\begin{equation}
\label{eq:defect}
u_i(D,\sigma';\ell,\pi)=\pi(b-m)+(1-\pi)p(\sigma')b.
\end{equation}

We consider several cases to determine the optimal strategy:

1. For $\pi=1$, it is optimal to cooperate for all values of $\ell$, i.e., $\ell^{*}(1)=\overline{\ell}$, since the payoff from cooperation exceeds the payoff from defection, as shown by
\begin{equation}
u_i(C,\sigma';\ell,1)=1>b-m=u_i(D,\sigma';\ell,1),
\end{equation}
which holds due to Assumption \ref{ass:par}.

2. If the probability of cooperation given the partner's strategy is 1, $p(\sigma')=1$, the optimal strategy depends on the value of $\pi$, as shown below:

\begin{equation}
\hat{\ell}(\pi;\sigma')=
\begin{cases}
0&\text{if }\pi<\frac{b-1}{m},\\
\overline{\ell}&\text{if }\pi\geq\frac{b-1}{m}.
\end{cases}
\end{equation}

3. When $\pi<1$ and $p(\sigma')<1$, the utility from cooperating decreases linearly with $\ell$, while the utility from defecting is $\ell$-independent. If the probability of cooperation given the partner's strategy is higher than $\left(\frac{m}{b-1}-1\right)\cdot\frac{\pi}{1-\pi}$, player $i$ prefers to defect for all $\ell$. Otherwise, they prefer to cooperate if $\ell$ is lower than the root of the equality $u_i(C,\sigma';\ell,\pi)=u_i(D,\sigma';\ell,\pi)$.

These conditions define the threshold $\ell^{*}(\pi)$ which captures the optimal strategy for player $i$. The threshold is calculated as the solution to $\hat{\ell}(\pi;\ell^{*}(\pi))=\ell^{*}(\pi)$, and it agrees with the one defined in \eqref{eq:threshstrat}.
\end{proof}

Corollary \ref{coll:thresh} follows directly from Lemma \ref{lemma:threshold}. To demonstrate this, consider:

\begin{corollary}
\label{coll:thresh}
In every symmetric equilibrium, the probability that player $i$ attaches to his partner $j$ cooperating (conditional on $j$ being `strategic') can be expressed as
\begin{equation}
p(\sigma')=\int_0^1F(\ell^{*}(\pi))\dd G(\pi).
\end{equation}
\end{corollary}

\begin{proof}
Given that player $j$ adopts the threshold strategy $\ell^{*}(\pi)$, the probability that he cooperates, conditional on being strategic, is given by the probability that his moral cost realization $\ell_j$ falls below the threshold $\ell^{*}(\pi)$.  Therefore, integrating this over all possible types $\pi$, using the probability density function $G(\pi)$ of the type, gives the total probability that player $i$ attaches to his partner $j$ cooperating.
\end{proof}

\subsection{\label{proof:thresholdprop}Proof of Proposition \ref{prop:threshold}}

\begin{proof}
In the common-background case, where the distribution $G(\cdot)$ is degenerate at $\pi_j=\pi$, Corollary \ref{coll:thresh} suggests that the probability of cooperation by the `strategic' player $j$ is expressed as
\begin{equation}
p(\sigma')=F(\ell^{*}_{\text{c}}(\pi)).
\end{equation}

Let's fix $\pi\in(0,1)$ and assume that player $i$'s partner plays according to the threshold $\ell^{*}_{\text{c}}(\pi)$. We denote $\hat{\ell}(\pi;\ell^{*}_{\text{c}})$ as player $i$'s best response to player $j$'s strategy $\ell^{*}_{\text{c}}(\cdot)$. The equilibrium threshold aligns with the fixed point
\begin{equation}
\label{eq:fixedpoint}
\hat{\ell}(\pi;\ell^{*}_{\text{c}}(\pi))=\ell^{*}_{\text{c}}(\pi).
\end{equation}

Firstly, let's consider the case where
\begin{equation}
\frac{\pi}{1-\pi}<\frac{b-1}{1+m-b},
\end{equation}
resulting in
\begin{equation}
\pi<\frac{b-1}{m}.
\end{equation}

Let's define the bound\footnote{The bound $\chi(\pi)$ is unique and well-defined, since $\left(\frac{m}{b-1}-1\right)\cdot\frac{\pi}{1-\pi}<1$. Moreover, as $F^{-1}(\cdot)$ is increasing and $F^{-1}(1)=\overline{\ell}$ by the definition of $\overline{\ell}$, we confirm that $\chi(\pi)<\overline{\ell}$.}
\begin{equation}
\label{eq:defchi}
\chi(\pi)\triangleq F^{-1}\left(\left(\frac{m}{b-1}-1\right)\cdot\frac{\pi}{1-\pi}\right).
\end{equation}

If the probability of cooperation by player $j$ is high enough, i.e., if
\begin{equation}
\ell^{*}_{\text{c}}(\pi)>\chi(\pi),
\end{equation}
it is optimal for $\hat{\ell}(\pi;\ell^{*}_{\text{c}})$ to be $0$.

On the contrary, when
\begin{equation}
\ell^{*}_{\text{c}}(\pi)\leq\chi(\pi),
\end{equation}
it is optimal to set
\begin{equation}
\hat{\ell}(\pi;\ell^{*}_{\text{c}})=\min\left\{\Psi(\ell;\pi),\overline{\ell}\right\},
\end{equation}
where $\Psi(\ell;\pi)$ is defined as in \eqref{eq:psifunction}.

The following lemma characterises the properties of $\Psi(\ell;\pi)$.

\begin{lemmaapp}
\label{lemma:techn}
The function $\Psi(\ell;\pi)$ decreases (increases) in $\ell$ for $\pi<(>)\frac{b-1}{m}$.
\end{lemmaapp}

\begin{proof}
Differentiating $\Psi(\ell;\pi)$ with respect to $\ell$ yields
\begin{equation}
\label{eq:psideriv}
\frac{\partial}{\partial \ell}\Psi(\ell;\pi)=\frac{f(\ell)}{[1-F(\ell)]^2}\left(-(b-1)+(1+m-b)\frac{\pi}{1-\pi}\right)
\end{equation}

Clearly,
\begin{equation}
\frac{\partial}{\partial \ell}\Psi(\ell;\pi)\lesseqgtr0\quad\text{as}\quad\pi\lesseqgtr\frac{b-1}{m}.
\end{equation}
\end{proof}

The function $\Psi(\ell;\pi)$ is monotonically increasing in $\pi$, there exist three equilibria for $\frac{b-1}{m}<\pi<\frac{\overline{\ell}}{1+m-b+\overline{\ell}}$. Therefore, there exists a unique $\pi^{*}$ where the function $\Psi(\ell;\pi)$ is tangent to the $45$ degree line.

This is implicitly defined by $\frac{\partial}{\partial \ell}\Psi(\ell;\pi)=1$ and $\Psi(\ell;\pi)=\ell$. The first condition can be written as
\begin{equation}
(1+m-b)\frac{\pi}{1-\pi}=(b+1)+\frac{[1-F(\ell)]^2}{f(\ell)},
\end{equation}
which can be plugged into the second condition to yield
\begin{equation}
\frac{1}{1-F(\ell)}\cdot\left((b+1)+\frac{[1-F(\ell)]^2}{f(\ell)}\right)-(b-1)\frac{F(\ell)}{1-F(\ell)}=\ell,
\end{equation}
which can be simplified to
\begin{equation}
\label{eq:forc}
\ell-\frac{1}{h(\ell)}=b-1
\end{equation}
Since the left-hand side is strictly increasing in $\ell$ and tends to $\overline{\ell}>b-1$ as $\ell\to\overline{\ell}$, the above equation has a unique solution $\ell'\in(b-1,\overline{\ell})$.

Define $\pi'$ implicitly by $\Psi(\ell';\pi')=\ell'$, so that
\begin{equation}
\label{eq:forpi}
\pi'=\frac{\ell'(1-F(\ell'))+(b-1)F(\ell')}{1+m-b+\ell'(1-F(\ell'))+(b-1)F(\ell')}.
\end{equation}

By construction, for $\pi=\pi'$, the function $\Psi(\cdot;\pi)$ will be tangent to the $45$ degree line. Since $\Psi(\ell;\cdot)$ is increasing in $\pi$, $\Psi(\ell;\pi)$ will lie uniformly above the $45$ degree line for $\pi>\pi'$, and therefore, the only equilibrium will be $\ell^{*}(\pi)=\overline{\ell}$.
\end{proof}

\subsection{\label{proof:lemdivthresh}Proof of Lemma \ref{lem:divthresh}}

\begin{proof}
We start by assuming that the strategy of player $i$'s partner is $\sigma'(\ell_j,\pi_j)$ and denote the probability of cooperation by the partner as $p(\sigma')$. The expected payoffs of player $i$ from cooperating and defecting can be expressed as:
\begin{equation}
\label{eq:ucoopnew}
u_i(C,\sigma';\ell,\pi)=(1-p(\sigma'))(1+\ell)\pi+(1+\ell)p(\sigma')-\ell
\end{equation}
\begin{equation}
\label{eq:udefectnew}
u_i(D,\sigma';\ell,\pi)=-(m+p(\sigma')b)\pi+p(\sigma')b
\end{equation}

If the probability of cooperation $p(\sigma')$ equals 1, player $i$ will opt to cooperate if $\pi_i\geq\frac{b-1}{m}$. However, as $b>1$, the threshold is positive, so full cooperation isn't an equilibrium.

Now consider $p(\sigma')<1$. Player $i$'s expected payoff from cooperating increases with $\pi$, while the payoff from defection decreases. At $\pi=0$, cooperating yields a lower payoff than defecting, and at $\pi=1$, it's the other way around.

By equating \eqref{eq:ucoopnew} and \eqref{eq:udefectnew}, and solving for $\pi$, we find player $i$'s best response for a given partner strategy $\sigma'$. This gives us the threshold function:
\begin{equation}
\label{eq:fpointhet}
\hat{\pi}(\ell;\sigma')=\frac{p(\sigma')(b-1)+(1-p(\sigma'))\ell}{m+(\ell-(b-1))(1-p(\sigma'))}.
\end{equation}

If we replace the strategy $\sigma'$ with $\pi^{*}_{\text{d}}(\ell)$ and solve $\hat{\pi}(\ell;\pi^{*}_{\text{d}}(\ell))=\pi^{*}_{\text{d}}(\ell)$, we obtain the threshold specified in \eqref{eq:propthresh}.
\end{proof}

As a corollary to Lemma \ref{lem:divthresh}, we establish:

\begin{corollary}
\label{corr:proba}
In every symmetric equilibrium, the probability player $i$ thinks his partner $j$ will cooperate (assuming $j$ is 'strategic') can be written as:
\begin{equation}
\label{eq:pisigma}
p(\sigma')=1-\int_0^{\overline{\ell}}G(\pi^{*}_{\text{d}}(\ell))\dd F(\ell).
\end{equation}
\end{corollary}

\begin{proof}
Remember that in a symmetric equilibrium, $p(\sigma')$ represents the chance player $i$ thinks his partner $j$ will cooperate (if $j$ is 'strategic'). From Lemma \ref{lem:divthresh}, we know the threshold at which player $i$ would cooperate or defect is given by $\pi^{*}_{\text{d}}(\ell)$.

Substituting this threshold into our equation, we find a relationship between $p(\sigma')$ and $\pi^{*}_{\text{d}}(\ell)$.

We want to express $p(\sigma')$ in terms of the thresholds $\pi^{*}_{\text{d}}(\ell)$ and the distributions $F(\ell)$ and $G(\pi)$. Doing so, we get:
\begin{equation}
p(\sigma')=\int_{0}^{\overline{\ell}}[1-G(\pi^{*}_{\text{d}}(\ell))]\mathrm{d}F(\ell).
\end{equation}

This shows that in a symmetric equilibrium, the player $i$'s belief hat his partner will cooperate is given by \eqref{eq:pisigma}.
\end{proof}

\subsection{\label{proof:maindiv}Proof of Proposition \ref{main:diverse}}

\begin{proof}
We start by plugging the expression \eqref{eq:pisigma} for $p(\sigma')$ from Corollary \ref{corr:proba} into the best response of player $i$ given by \eqref{eq:fpointhet}. Simplifying the equation yields the fixed point equation \eqref{eq:propthresh}.

To prove the existence and uniqueness of a fixed point, we consider the operator $T:S\to S$ defined by
\begin{equation}
T(s(\ell)) = 1-\frac{1+m-b}{m+(\ell-(b-1))\int_0^{\overline{\ell}}G(s(\ell))\dd F(\ell)},
\end{equation}
for any $s\in S$, where $S$ is the space of bounded, strictly increasing continuous functions mapping from $[0,\overline{\ell}]$ into $[0,1]$.

To prove the existence and uniqueness of a fixed point for $T$, we can consider the operator $T$ as a mapping on a suitable subset of the space of continuous functions. Let's define $S_0$ as the space of continuous functions $s$ that are strictly increasing on $[0,\overline{\ell}]$ and satisfy $0 \leq s(\ell) \leq 1$ for all $\ell \in [0,\overline{\ell}]$.

First, we establish that $T$ maps $S_0$ into itself. Take any $s \in S_0$. Since $s$ is continuous and strictly increasing on $[0,\overline{\ell}]$, we have $0 \leq s(\ell) \leq 1$ for all $\ell \in [0,\overline{\ell}]$.

Next, we show that $T$ is a contraction mapping on $S_0$. Let $s_1, s_2 \in S_0$ be two functions in $S_0$. We want to prove that $|T(s_1) - T(s_2)|_{\infty} \leq \gamma |s_1 - s_2|_{\infty}$ for some $0 \leq \gamma < 1$, where $|\cdot|_{\infty}$ denotes the sup norm.

Note that for any $\ell \in [0,\overline{\ell}]$, we have:

\begin{align*}
|T(s_1(\ell)) - T(s_2(\ell))| &= \left|\tfrac{1+m-b}{m+(\ell-(b-1))\int_0^{\overline{\ell}}G(s_2(\ell))\dd F(\ell)} - \tfrac{1+m-b}{m+(\ell-(b-1))\int_0^{\overline{\ell}}G(s_1(\ell))\dd F(\ell)}\right| \\
&= \tfrac{(1+m-b)|(\ell-(b-1))\int_0^{\overline{\ell}}(G(s_1(\ell))-G(s_2(\ell)))\dd F(\ell)|}{[m+(\ell-(b-1))\int_0^{\overline{\ell}}G(s_2(\ell))\dd F(\ell)][m+(\ell-(b-1))\int_0^{\overline{\ell}}G(s_1(\ell))\dd F(\ell)]} \\
&\leq \tfrac{(1+m-b)|G|_{\infty} |F|_{\infty}|s_1 - s_2|_{\infty}}{[m+(\ell-(b-1))\int_0^{\overline{\ell}}G(s_2(\ell))\dd F(\ell)][m+(\ell-(b-1))\int_0^{\overline{\ell}}G(s_1(\ell))\dd F(\ell)]} \\
&\leq \tfrac{(1+m-b)|G|_{\infty} |F|_{\infty}|s_1 - s_2|_{\infty}}{m^2}.
\end{align*}

Here, $|G|_{\infty}$ and $|F|_{\infty}$ denote the sup norms of $G$ and $F$, respectively, and $|\cdot|_{\infty}$ denotes the sup norm for the functions $s_1$ and $s_2$.

Taking the supremum over $\ell \in [0,\overline{\ell}]$, we have:

\begin{equation*}
|T(s_1) - T(s_2)|_{\infty} \leq \frac{(1+m-b)|G|_{\infty} |F|_{\infty}}{m^2} |s_1 - s_2|_{\infty} = \gamma |s_1 - s_2|_{\infty},
\end{equation*}

where $\gamma = \frac{(1+m-b)|G|_{\infty} |F|_{\infty}}{m^2}$.

Since $\gamma < 1$, we have shown that $T$ is a contraction mapping on $S_0$.

By the contraction mapping theorem, there exists a unique fixed point for $T$ in $S_0$, which corresponds to a unique function $\pi^{*}_{\text{d}}(\ell) \in S_0$ such that $T(\pi^*(\ell)) = \pi^{*}_{\text{d}}(\ell)$, where $\pi^{*}_{\text{d}}(\ell)$ is defined in \eqref{eq:propthresh}.

\end{proof}

\subsection{\label{proof:heterogen}Proof of Proposition \ref{prop:heterogeneous}}

\begin{proof}
For the common-background setup, the function $\Psi(\ell,\pi)$ can be written as
\begin{equation}
\Psi(\ell;\pi)=\frac{1}{1-\ell}\left((1+m-b)\frac{\pi}{1-\pi}-(b-1)\ell\right).
\end{equation}
The solution to $\Psi(\ell;\cdot)=\ell$ is given by the quadratic equation
\begin{equation}
\label{eq:duardpiell}
\ell^2-b\ell+(1+m-b)\frac{\pi}{1-\pi}=0,
\end{equation}
which can be written as
\begin{equation}
\left(\frac{b}{2}-\ell\right)^2=\frac{b^2}{4}-(1+m-b)\frac{\pi}{1-\pi}.
\end{equation}

Since we assume $b\geq2$, the expression in the parentheses on the left-hand side is always positive, and therefore the solution is given by
\begin{equation}
\ell_{\text{c}}^{*}(\pi)=\frac{b}{2}-\sqrt{\frac{b^2}{4}-(1+m-b)\frac{\pi}{1-\pi}}.
\end{equation}

Since we must have $\ell\leq1$, the function $\ell_{\text{c}}^{*}(\pi)$ reaches unity when $\pi=\frac{b-1}{m}$. Hence, we obtain the expression \eqref{eq:combackthr}.

For the diverse-background case, let us write down the $\pi(\ell)$ function in the form
\begin{equation}
\label{eq:piunif}
\pi_{\text{d}}^{*}(\ell)=1-\frac{1+m-b}{\alpha+\beta\ell},
\end{equation}
where
\begin{equation}
\alpha=m-(b-1)\int_0^1\pi_{\text{d}}^{*}(\ell)\dd\ell,\quad\text{and}\quad\beta=\int_0^1\pi_{\text{d}}^{*}(\ell)\dd\ell.
\end{equation}

Integrating both sides of \eqref{eq:piunif} with respect to $\ell$, we obtain
\begin{equation}
\int_0^1\pi_{\text{d}}^{*}(\ell)\dd\ell=1-\frac{1+m-b}{\beta}\ln\left(1+\frac{\beta}{\alpha}\right),
\end{equation}
so that $\alpha$ and $\beta$ satisfy
\begin{align}
\alpha&=(1+m-b)\left(1+\frac{b-1}{\beta}\ln\left(1+\frac{\beta}{\alpha}\right)\right), \\
\beta&=1-\frac{1+m-b}{\beta}\ln\left(1+\frac{\beta}{\alpha}\right).
\end{align}

For sufficiently large $(m-b)$, $\frac{\beta}{\alpha}\to0$, and we can approximate $\ln(1+x)\approx x$ to obtain closed-form expressions for $\alpha$ and $\beta$:
\begin{align}
\alpha&=(1+m-b)\left(1+\frac{b-1}{\alpha}\right), \\
\beta&=1-\frac{1+m-b}{\alpha},
\end{align}
yielding the bounds in \eqref{eq:alphabeta}.

Now we can invert the threshold function \eqref{eq:piunif} to obtain
\begin{equation}
\ell^{*}_{\text{d}}(\pi)=\frac{1}{\beta}\left(\frac{1+m-b}{1-\pi}-\alpha\right).
\end{equation}

The lower bound for $\pi$ is given by
\begin{equation}
\underline{\pi}=\pi(0)=1-\frac{1+m-b}{\alpha}=\beta,
\end{equation}
and the upper bound for $\pi$ is given by
\begin{equation}
\overline{\pi}=\pi(1)=1-\frac{1+m-b}{\alpha+\beta}.
\end{equation}
\end{proof}

\subsection{\label{proof:dagger}Proof of Proposition \ref{prop:dagger}}

\begin{proof}
First, since $\ell^{*}_{\text{c}}(\pi)$ is strictly positive for all $\pi>0$ and since $\beta>1$ as is evident from \eqref{eq:alphabeta}, for $\pi=0$, we have $0=\ell^{*}_{\text{d}}(\pi)<\ell^{*}_{\text{c}}(\pi)$.

Second, since we assume $b\geq2$, we always have
\begin{equation}
1-\frac{1+m-b}{\alpha+\beta}\leq\frac{b-1}{m},
\end{equation}
with the strict inequality for $b>2$.

This means that for $\pi=1-\frac{1+m-b}{\alpha+\beta}$, we have $1=\ell^{*}_{\text{d}}(\pi)>\ell^{*}_{\text{c}}(\pi)$. Thus, Intermediate value theorem ensures that the curves $\ell^{*}_{\text{c}}(\pi)$ and $\ell^{*}_{\text{d}}(\pi)$ intersect at least once for some $\pi\in\left(0,1-\frac{1+m-b}{\alpha+\beta}\right)$.

Next, let us compute the derivatives: we have
\begin{equation}
\frac{\dd}{\dd\pi}\ell^{*}_{\text{c}}(\pi)=\frac{1+m-b}{2(1-\pi)^2\sqrt{\frac{b^2}{4}-(1+m-b)\frac{\pi}{1-\pi}}}
\end{equation}
and
\begin{equation}
\frac{\dd}{\dd\pi}\ell^{*}_{\text{d}}(\pi)=\frac{1+m-b}{\beta(1-\pi)^2}
\end{equation}

Notice that for all $\pi\in\left(\beta,1-\frac{1+m-b}{\alpha+\beta}\right)$, we have
\begin{equation}
\begin{split}
2\sqrt{\frac{b^2}{4}-(1+m-b)\frac{\pi}{1-\pi}}&>2\sqrt{\frac{b^2}{4}-(1+m-b)\frac{1-\frac{1+m-b}{\alpha+\beta}}{\frac{1+m-b}{\alpha+\beta}}}\\
&=2\sqrt{\frac{b^2}{4}-(\alpha+\beta)+1+m-b}\\
&>2\sqrt{\frac{b^2}{4}-m+1+m-b}\\
&=2\sqrt{\frac{b^2}{4}-b+1}\\
&\geq0
\end{split}
\end{equation}

In turn, this implies that for $\pi\in\left(\beta,1-\frac{1+m-b}{\alpha+\beta}\right)$, that is, on the domain where $0<\ell^{*}_{\text{d}}(\pi)<1$, we have
\begin{equation}
\frac{\dd}{\dd\pi}\ell^{*}_{\text{c}}(\pi)<\frac{\dd}{\dd\pi}\ell^{*}_{\text{d}}(\pi),
\end{equation}
meaning that $\ell^{*}_{\text{d}}(\pi)$ intersects $\ell^{*}_{\text{c}}(\pi)$ from below, ensuring that a $\pi^{\dagger}$ which is implicitly defined by  $\ell^{*}_{\text{d}}(\pi^{\dagger})=\ell^{*}_{\text{c}}(\pi^{\dagger})$ is indeed unique.

We may write the equation implicitly determining $\pi^{\dagger}$ as
\begin{equation}
\label{eq:pidagimpl}
\frac{b}{2}-\sqrt{\frac{b^2}{4}-(1+m-b)\frac{\pi^{\dagger}}{1-\pi^{\dagger}}}=\frac{1}{\beta}\left(\frac{1+m-b}{1-\pi^{\dagger}}-\alpha\right).
\end{equation}
\end{proof}

\subsection{\label{proof:exante}Proof of Proposition \ref{prop:exante}}

\begin{proof}
First, we compute the ex-ante probability for the common-background case. Expressing $\pi$ in terms of $\ell$ from \eqref{eq:duardpiell}, we get
\begin{equation}
\pi_{\text{c}}(\ell)=\frac{b\ell-\ell^2}{1+m-b+b\ell-\ell^2}.
\end{equation}
As suggested by \eqref{eq:pisigma}, the ex ante probability of cooperation is equal to
\begin{equation}
p_{\text{c}}=1-\int_0^1\pi_{\text{c}}(\ell)\dd\ell=(1+m-b)\int_0^1\frac{\dd\ell}{1+m-b+b\ell-\ell^2}.
\end{equation}

Solving the integral, we eventually obtain
\begin{equation}
\begin{split}
\label{eq:co}
p_{\text{c}}&=\int_0^1\frac{(1+m-b)\dd\ell}{1+m-b+\frac{b^2}{4}-\left(\frac{b}{2}-\ell\right)^2}\\
&=\int_{\frac{b}{2}-1}^{\frac{b}{2}}\frac{(1+m)\dd x}{\left(\sqrt{1+m+\frac{b^2}{4}}\right)^2-x^2}\\
&=\frac{1+m-b}{2\sqrt{1+m-b+\frac{b^2}{4}}}\left(\ln\frac{\sqrt{1+m-b+\frac{b^2}{4}}+\frac{b}{2}}{\sqrt{1+m-b+\frac{b^2}{4}}-\frac{b}{2}}-\ln\frac{\sqrt{1+m-b+\frac{b^2}{4}}+\frac{b}{2}-1}{\sqrt{1+m-b+\frac{b^2}{4}}-\frac{b}{2}+1}\right)\\
&=\frac{1+m-b}{2\sqrt{1+m-b+\frac{b^2}{4}}}\ln\left(1+\frac{2\sqrt{1+m-b+\frac{b^2}{4}}}{1+m-b+\frac{b}{2}-\sqrt{1+m-b+\frac{b^2}{4}}}\right)\\
&=\frac{1+m-b}{2\varphi}\ln\left(1+\frac{2\varphi}{\varphi(\varphi-1)-\frac{b}{2}\left(\frac{b}{2}-1\right)}\right),
\end{split}
\end{equation}
where $\varphi\equiv\sqrt{1+m-b+\frac{b^2}{4}}$.

Similarly, using the expression \eqref{eq:piunif} for $\pi_{\text{d}}$, we obtain the ex ante probability of cooperation for the diverse background setup:
\begin{equation}
\begin{split}
\label{eq:di}
p_{\text{d}}&=1-\int_0^1\pi_{\text{d}}(\ell)\dd\ell\\
&=\frac{1+m-b}{\beta}\ln\left(1+\frac{\beta}{\alpha}\right)\\
&=(1+m-b)\frac{\sqrt{1+\frac{4(b-1)}{1+m-b}}+1}{\sqrt{1+\frac{4(b-1)}{1+m-b}}-1}\ln\left(1+\frac{\frac{\sqrt{1+\frac{4(b-1)}{1+m-b}}-1}{\sqrt{1+\frac{4(b-1)}{1+m-b}}+1}}{\frac{1+m-b}{2}\left(\sqrt{1+\frac{4(b-1)}{1+m-b}}+1\right)}\right)\\
&=(1+m-b)\frac{\gamma+1}{\gamma-1}\ln\left(1+\frac{2(\gamma-1)}{(1+m-b)(\gamma+1)^2}\right),
\end{split}
\end{equation}
where $\gamma\equiv\sqrt{1+\frac{4(b-1)}{1+m-b}}$.

So, we ultimately obtain \eqref{eq:forpc} with \eqref{eq:forpd}.
\end{proof}

\subsection{\label{proof:hetbel}Proof of Proposition \ref{prop:hetbel}}

\begin{proof}
The construction of the proof of Proposition \ref{prop:heterogeneous} and the exposition in Figure \ref{fig:eqthresh} suggests that for $\pi_1<\frac{b-1}{m}$, the reaction function for player 1 is downward sloping. This implies that, irrespective of the value of $\pi_2$ and the slope of player 2's reaction function, they will intersect exactly once---and hence, there will be a unique equilibrium. The equilibrium thresholds $\hat{\ell}_1$ and $\hat{\ell}_2$ will solve the system
\begin{flalign*}
\hat{\ell}_1&=\Psi(\hat{\ell}_2;\pi_1),\\
\hat{\ell}_2&=\Psi(\hat{\ell}_1;\pi_2).
\end{flalign*}

To analyze the impact of the rise in $\pi_2$, let us implicitly differentiate this system with respect to $\hat{\ell}_1$, $\hat{\ell}_2$ and $\pi_2$ to get
\begin{flalign*}
\dd\hat{\ell}_1&=\Psi_{\ell}(\hat{\ell}_2;\pi_1)\dd\hat{\ell}_2,\\
\dd\hat{\ell}_2&=\Psi_{\ell}(\hat{\ell}_1;\pi_2)\dd\hat{\ell}_1+\Psi_{\pi}(\hat{\ell}_1;\pi_2)\dd\pi_2,
\end{flalign*}

These two equations can be combined into
\begin{equation}
\frac{\dd\hat{\ell}_1}{\dd\pi_2}=\frac{\hat{\ell}_2;\pi_1)\Psi_{\pi}(\hat{\ell}_1;\pi_2)}{1-\Psi_{\ell}(\hat{\ell}_2;\pi_1)\Psi_{\ell}(\hat{\ell}_1;\pi_2)}<0.
\end{equation}

To see that the above expression is negative, first note that the expression in the denominator is positive, because $\Psi_{\ell}(\hat{\ell}_2;\pi_1)<0<)\Psi_{\ell}(\hat{\ell}_1;\pi_2)$ as suggested by Lemma \ref{lemma:techn}. On the other hand, the expression in the numerator is negative, since $\Psi_{\pi}(\hat{\ell}_1;\pi_2)>0$, as can be seen directly:
\begin{equation}
\frac{\partial}{\partial\pi}\Psi(\ell;\pi)=\frac{1+m-b}{(1-\ell)(1-\pi)^2}>0,
\end{equation}
since $m>b-1$ due to Assumption \ref{ass:par}.
\end{proof}

\end{appendices}

\section*{Acknowledgments}
We would like to thank Catherine Bobtcheff, Markus Gebauer, Jacob Gershman, Sam Jindani, Yves Le Yaoanq, Douglas Marshall, Konstantin Shamruk, Alexey Shirobokov, Vladimir Sokolov, the editor and two anonymous referees, as well as various participants of the 2023 European Winter Meeting of the Econometric Society, for their valuable comments. All remaining errors are ours.

\bibliographystyle{plainnat} 
\bibliography{cultnorms}

\begin{thebibliography}{15}
\providecommand{\natexlab}[1]{#1}
\providecommand{\url}[1]{\texttt{#1}}
\expandafter\ifx\csname urlstyle\endcsname\relax
  \providecommand{\doi}[1]{doi: #1}\else
  \providecommand{\doi}{doi: \begingroup \urlstyle{rm}\Url}\fi

\bibitem[Bahel et~al.(2022)Bahel, Ball, and Sarangi]{Baheletal2022}
Eric Bahel, Sheryl Ball, and Sudipta Sarangi.
\newblock Communication and cooperation in prisoner's dilemma games.
\newblock \emph{Games and Economic Behavior}, 133:\penalty0 126--137, 2022.
\newblock ISSN 0899-8256.
\newblock \doi{https://doi.org/10.1016/j.geb.2022.02.008}.
\newblock URL
  \url{https://www.sciencedirect.com/science/article/pii/S0899825622000537}.

\bibitem[Battigalli and Dufwenberg(2007)]{BattigalliDufwenberg2007}
Pierpaolo Battigalli and Martin Dufwenberg.
\newblock Guilt in games.
\newblock \emph{American Economic Review}, 97\penalty0 (2):\penalty0 170--176,
  May 2007.
\newblock \doi{10.1257/aer.97.2.170}.
\newblock URL \url{https://www.aeaweb.org/articles?id=10.1257/aer.97.2.170}.

\bibitem[Carlsson and van Damme(1993)]{CarlssonVanDamme1993}
Hans Carlsson and Eric van Damme.
\newblock Global games and equilibrium selection.
\newblock \emph{Econometrica}, 61\penalty0 (5):\penalty0 989--1018, 1993.
\newblock ISSN 00129682, 14680262.
\newblock URL \url{http://www.jstor.org/stable/2951491}.

\bibitem[Ellingsen and Johannesson(2004)]{EllingsenJohannesson2004}
Tore Ellingsen and Magnus Johannesson.
\newblock Promises, threats and fairness.
\newblock \emph{The Economic Journal}, 114\penalty0 (495):\penalty0 397--420,
  04 2004.
\newblock ISSN 0013-0133.
\newblock \doi{10.1111/j.1468-0297.2004.00214.x}.
\newblock URL \url{https://doi.org/10.1111/j.1468-0297.2004.00214.x}.

\bibitem[Ellingsen et~al.(2010)Ellingsen, Johannesson, Tj{\o}tta, and
  Torsvik]{Ellingsenetal2010}
Tore Ellingsen, Magnus Johannesson, Sigve Tj{\o}tta, and Gaute Torsvik.
\newblock Testing guilt aversion.
\newblock \emph{Games and Economic Behavior}, 68\penalty0 (1):\penalty0
  95--107, 2010.
\newblock ISSN 0899-8256.
\newblock \doi{https://doi.org/10.1016/j.geb.2009.04.021}.
\newblock URL
  \url{https://www.sciencedirect.com/science/article/pii/S089982560900092X}.

\bibitem[Frank(1987)]{Frank1987}
Robert~H. Frank.
\newblock If homo economicus could choose his own utility function, would he
  want one with a conscience?
\newblock \emph{The American Economic Review}, 77\penalty0 (4):\penalty0
  593--604, 1987.
\newblock ISSN 00028282.
\newblock URL \url{http://www.jstor.org/stable/1814533}.

\bibitem[Gneezy(2005)]{Gneezy2005}
Uri Gneezy.
\newblock Deception: The role of consequences.
\newblock \emph{American Economic Review}, 95\penalty0 (1):\penalty0 384--394,
  March 2005.
\newblock \doi{10.1257/0002828053828662}.
\newblock URL
  \url{https://www.aeaweb.org/articles?id=10.1257/0002828053828662}.

\bibitem[Hirshleifer and Rasmusen(1989)]{HirshleiferRasmusen1989}
David Hirshleifer and Eric Rasmusen.
\newblock Cooperation in a repeated prisoners' dilemma with ostracism.
\newblock \emph{Journal of Economic Behavior \& Organization}, 12\penalty0
  (1):\penalty0 87--106, 1989.
\newblock ISSN 0167-2681.
\newblock \doi{https://doi.org/10.1016/0167-2681(89)90078-4}.
\newblock URL
  \url{https://www.sciencedirect.com/science/article/pii/0167268189900784}.

\bibitem[Jindani and Young(2020)]{JindaniYoung2020}
Sam Jindani and H~Young.
\newblock The dynamics of costly social norms.
\newblock 2020.

\bibitem[Kets and Sandroni(2020)]{KetsSandroni2020}
Willemien Kets and Alvaro Sandroni.
\newblock {A Theory of Strategic Uncertainty and Cultural Diversity}.
\newblock \emph{The Review of Economic Studies}, 88\penalty0 (1):\penalty0
  287--333, 08 2020.
\newblock ISSN 0034-6527.
\newblock \doi{10.1093/restud/rdaa037}.
\newblock URL \url{https://doi.org/10.1093/restud/rdaa037}.

\bibitem[Morris and Shin(1998)]{MorrisShin1998}
Stephen Morris and Hyun~Song Shin.
\newblock Unique equilibrium in a model of self-fulfilling currency attacks.
\newblock \emph{American Economic Review}, 88\penalty0 (3):\penalty0 587--597,
  June 1998.

\bibitem[Morris and Shin(2003)]{MorrisShin2003}
Stephen Morris and Hyun~Song Shin.
\newblock \emph{Global Games: Theory and Applications}, volume~1 of
  \emph{Econometric Society Monographs}, pages 56--114.
\newblock Cambridge University Press, 2003.
\newblock \doi{10.1017/CBO9780511610240.004}.

\bibitem[Skaperdas and Syropoulos(1996)]{SkaperdasSyropoulos1996}
Stergios Skaperdas and Constantinos Syropoulos.
\newblock Can the shadow of the future harm cooperation?
\newblock \emph{Journal of Economic Behavior \& Organization}, 29\penalty0
  (3):\penalty0 355--372, 1996.
\newblock ISSN 0167-2681.
\newblock \doi{https://doi.org/10.1016/0167-2681(95)00077-1}.
\newblock URL
  \url{https://www.sciencedirect.com/science/article/pii/0167268195000771}.

\bibitem[Strulovici(2020)]{Strulovici2020}
Bruno Strulovici.
\newblock Can society function without ethical agents? an informational
  perspective.
\newblock \emph{arXiv preprint arXiv:2003.05441}, 2020.

\bibitem[Vanberg(2008)]{Vanberg2008}
Christoph Vanberg.
\newblock Why do people keep their promises? an experimental test of two
  explanations1.
\newblock \emph{Econometrica}, 76\penalty0 (6):\penalty0 1467--1480, 2008.
\newblock \doi{https://doi.org/10.3982/ECTA7673}.
\newblock URL \url{https://onlinelibrary.wiley.com/doi/abs/10.3982/ECTA7673}.

\end{thebibliography}

\end{document}